\documentclass[a4paper,oneside]{article}
\usepackage{amssymb}
\usepackage{graphicx}
\usepackage{multicol} 
\usepackage{floatflt}
\usepackage{color}
\definecolor{rosso}{cmyk}{0,1,1,0.4}
\definecolor{rossos}{cmyk}{0,1,1,0.55}
\definecolor{rossoc}{cmyk}{0,1,1,0.2}
\definecolor{blu}{cmyk}{1,1,0,0.3}
\definecolor{blus}{cmyk}{1,1,0,0.6}
\definecolor{bluc}{cmyk}{1,1,0,0.1}
\definecolor{verde}{cmyk}{0.92,0,0.59,0.25}
\definecolor{verdec}{cmyk}{0.92,0,0.59,0.15}
\definecolor{verdes}{cmyk}{0.92,0,0.59,0.4}

\advance\textheight by \topskip
    \newcommand{\cm}{{\rm cm}}

\newcommand{\meV}{\,{\rm meV}}

\large\normalsize
\newcommand{\be}{\begin{equation}}
\newcommand{\ee}{\end{equation}}
\newcommand{\ba}{\begin{array}}
\newcommand{\ea}{\end{array}}
\newcommand{\dm}{\Delta m^2}

\newcommand{\eV}{{\rm eV}}

\newcommand{\dchi}{$\Delta\chi^2$}
\def\Red  {}

\def\Black{}
\def\Blue {}
\newcommand{\eq}[1]{~(\ref{eq:#1})}
\newcommand{\MeV}{\,{\rm MeV}}
\newcommand{\NP}{Nucl. Phys.}
\newcommand{\PRL}{Phys. Rev. Lett.}
\newcommand{\PL}{Phys. Lett.}
\newcommand{\PR}{Phys. Rev.}
\newcommand{\mb}[1]{\mbox{\normalsize\boldmath $#1$}}

\def\circa#1{\,\raise.3ex\hbox{$#1$\kern-.75em\lower1ex\hbox{$\sim$}}\,}
\makeatletter
\def\art{\@ifnextchar[{\eart}{\oart}}
\def\eart[#1]#2#3#4#5#6{{\rm #2}, {\em #3 \bf #4} {\rm (#6) #5} ({\em #1})}
\def\hepart[#1]#2{{\rm #2, \em#1}}
\newcommand{\oart}[5]{{\rm #1}, {\em #2 \bf #3} {\rm (#5) #4}}

\newcounter{alphaequation}[equation]
\def\thealphaequation{\theequation\hbox to
0.6em{\hfil\alph{alphaequation}\hfil}}
\def\eqnsystem#1{
\def\@eqnnum{{\rm (\thealphaequation)}}
\def\@@eqncr{\let\@tempa\relax \ifcase\@eqcnt \def\@tempa{& & &} \or
  \def\@tempa{& &}\or \def\@tempa{&}\fi\@tempa
  \if@eqnsw\@eqnnum\refstepcounter{alphaequation}\fi
\global\@eqnswtrue\global\@eqcnt=0\cr}
\refstepcounter{equation} \let\@currentlabel\theequation \def\@tempb{#1}
\ifx\@tempb\empty\else\label{#1}\fi
\refstepcounter{alphaequation}
\let\@currentlabel\thealphaequation
\global\@eqnswtrue\global\@eqcnt=0 \tabskip\@centering\let\\=\@eqncr
$$\halign to \displaywidth\bgroup \@eqnsel\hskip\@centering
$\displaystyle\tabskip\z@{##}$&\global\@eqcnt\@ne
\hskip2\arraycolsep\hfil${##}$\hfil& \global\@eqcnt\tw@\hskip2\arraycolsep
$\displaystyle\tabskip\z@{##}$\hfil
\tabskip\@centering&\llap{##}\tabskip\z@\cr}
\def\endeqnsystem{\@@eqncr\egroup$$\global\@ignoretrue} \makeatother

\begin{document}%\twocolumn[
\centerline{hep-ph/0102234 \hfill CERN--TH/2001--46\hfill IFUP--TH/2001--6 \hfill SNS-PH/01--3}
\vspace{5mm}
\Black
\vspace{0.5cm}
\centerline{\LARGE\bf\Red Frequentist analyses of solar neutrino data}
\centerline{\large\bf(updated including the first CC data from SNO\footnote{The addendum at pages \pageref{6in}--\pageref{6out}
(section 6) is not present in the published version of this paper.})}
\centerline{\large\bf(updated including the NC and day/night data from SNO\footnote{The addendum at pages \pageref{7in}--\pageref{7out}
(section 7) is not present in the published version of this paper.})}
\centerline{\large\bf(updated including the first data from KamLAND\footnote{The addendum at pages \pageref{8in}--\pageref{8out}
(section 8) is not present in the published version of this paper.})}
\centerline{\large\bf(updated including SNO data with enhanced NC sensitivity\footnote{The addendum at pages \pageref{9in}--\pageref{9out}
(section 9) is not present in the published version of this paper.})}
\centerline{\large\bf(updated including 
final SNO `salt' data and 2004 KamLAND data\footnote{The addendum at page \pageref{10in}
(section 10) is not present in the published version of this paper.})}

\medskip\bigskip\Black
\centerline{\large\bf Paolo Creminelli, Giovanni Signorelli}\vspace{0.2cm}
\centerline{\em Scuola Normale Superiore and INFN, Sezione di Pisa, Italy}
\vspace{3mm}
\centerline{\large\bf Alessandro Strumia}\vspace{0.2cm}
\centerline{\em CERN, Geneva, Switzerland and Dipartimento di Fisica dell'Universit\`a di Pisa and INFN}
\vspace{1cm}
\Blue\centerline{\large\bf Abstract}
\begin{quote}\large\indent
The solar neutrino data are analysed in a frequentist framework, using the Crow--Gardner and 
Feldman--Cousins prescriptions for the construction of confidence regions. 
Including in the fit only the total rates measured by the various experiments, both methods give results similar to the 
commonly used $\Delta\chi^2$-cut approximation.
When fitting the full data set, the $\Delta\chi^2$-cut still gives a good approximation of the Feldman--Cousins regions.
However, a careful statistical analysis
significantly reduces the goodness-of-fit of the SMA and LOW solutions.
%   The different properties of these procedures emerge more clearly when we fit the full data set.
%   Finally, we clarify a significant issue concerning the goodness-of-fit.

In the addenda we discuss the implications of the latest
KamLAND, SNO and SK data.

\Black
\end{quote}

\section{Introduction}
The solar neutrino anomaly is an old but still controversial problem, in which many 
experimental data~\cite{ClSun,KaSun,GaSun,ExpsSun}  and
theoretical ingredients~\cite{BP98,MSW,Parke,BahcallWWW,LisiChiq}
have to be merged to give predictions for the oscillation parameters
and to rule out other non-standard explanation of the anomaly. 
A correct statistical treatment is a necessary step of the analysis. 
The starting point for interpreting the results of an experiment 
is the fact that one knows
the probability distribution $p({\rm data}|{\rm theory})$
for obtaining a set of data under the assumption that a given
theory is true. 
In the case of the solar neutrino anomaly (at least in its simplest version),
we know
$p(R_i | \Delta m^2,\theta)$, where 
$R_i$ are the three neutrino rates measured
in Chlorine, Gallium and SK experiments,
which should be used to infer the values of the theoretical
parameters $\Delta m^2$ and $\theta$.

This can be done according to
two conceptually very different approaches~\cite{StatInference},
each one with unsatisfactory aspects.
\begin{itemize}
\item {\bf The Bayesian approach} 
% considers probability as a tool that allows to quantify our level of knowdlege in conditions of uncertainty:
employs a probability distribution $p(\Delta m^2,\theta)$ to summarize our knowledge of
the parameters of the theory.
According to elementary properties of probability, this probability gets updated by the inclusion
of the results of a new experiment as
$p(\Delta m^2,\theta|R) \propto p(R|\Delta m^2,\theta) p(\Delta m^2,\theta)$.
The drawback is that one needs to choose some `prior' $p(\Delta m^2,\theta)$ to start with,
and the final result depends on this choice until experiments are sufficiently precise.
At the moment, solar neutrino fits give multiple distinct solutions so that
$p(\Delta m^2,\theta|R)$ still contains arbitrary order 1 factors.
The advantage is its extreme simplicity:
the laws of probability dictate what to do in any situation.

\item {\bf The frequentist approach} refuses the concept of probability of
theoretical parameters.
%% Despite this self mutilation
The Neyman construction~\cite{Neyman} allows us to
build range of parameters for any possible outcome of an experiment
with the property that $90\%$ (or whatever) of such ranges contain the true value.
However this procedure is not univocal and the resulting regions can be quite different. 
For example the Crow--Gardner~\cite{CG} procedure gives smaller regions  
in presence of unlikely statistical fluctuations in the measured outcome of the experiment, while
the Feldman--Cousins~\cite{FC} procedure gives ranges of roughly the same size for all possible outcomes.

%\footnote{One practical example:
%should we consider
%the uncertainties on the solar fluxes as statistical experimental unceratinties
%or as systematic uncertainties,
%or as theorethical uncertainties
%(that in a frequentist approach are not even allowed to exist)?
%The standard procedure consists in not raising such questions and
%summing all errors in quadrature.
%Indeed one can see that this is the right thing to do (in the Gaussian limit) 
%if one applies
%the Neyman construction in a Bayesian framework.}.

\end{itemize}
In simple cases when $p({\rm data}|{\rm
theory})$ is a Gaussian function of all its arguments (data and parameters, with no physical constraints on them), 
the Bayesian approach
(using a flat prior
$p$) and the frequentist approach (using the Feldman--Cousins method) 
are numerically equivalent to the commonly employed $\Delta\chi^2$-cut approximation.

When fitting solar neutrino data one has to be careful because:
\begin{itemize}
\item[(1)] $p(R_i | \Delta m^2,\theta)$ is a highly non-Gaussian function
of $\Delta m^2,\theta$:
in fact one finds a few separate
best-fit solutions (usually named `LMA', `SMA', `LOW', `VO')
while a Gaussian would have only one peak.
This is the problem that we will address in this paper.

\item[(2)] $p(R_i | \Delta m^2,\theta)$ is not perfectly
Gaussian as a function of $R_i$.
Assuming a Gaussian uncertainty on the detection cross 
sections $\sigma$ and on the solar fluxes $\Phi$,
one does not obtain a Gaussian uncertainty on the rates
$R\sim \sigma\cdot \Phi$.
In principle this is true; in practice 
the errors on $\sigma$ and $\Phi$
are sufficiently small that their product is also
almost Gaussian, up to very good accuracy.
%%%% perche'avevi levato la spiegazione?  Il motivo e'giusto ed ovvio.
%%%% Ogni funzione e'una somma, in approssimazione lineare.
\end{itemize}
Such issues have been studied in~\cite{giunti},
finding that (1) apparently has a dramatic effect (see fig.\ 3 of~\cite{giunti}: LMA and LOW merge in a single region), while
(2)  has a negligible effect
(see table II of~\cite{giunti}).
We will ignore (2) and
we therefore write the probability density function (pdf) for all the $n$ solar neutrino 
data $x_i$ as
\be
\label{eq:gaussian}
p(\mb{x}|\dm,\theta) = \frac{\exp[-\chi^2/2]}{(2 \pi)^{n/2} \sqrt{\det \sigma^2}} ,\qquad
\chi^2\equiv \sum_{i,j=1}^n (x_i^{\rm exp} - x_i^{\rm th})
\frac{1}{\sigma^{2}_{ij}} (x_j^{\rm exp} - x_j^{\rm th}).
\ee
The predicted values $x^{\rm th}$ and the covariance matrix $\sigma^2$ 
depend on $\dm$ and $\theta$. The covariance matrix contains both theoretical
and experimental errors, statistical and systematic, added in quadrature. 
This is the standard procedure, which can be justified by applying
the Neyman construction in a Bayesian framework
(i.e.\ by describing theoretical and systematic uncertainties using a probability distribution).
A strict frequentist framework employs a definition of probability that
makes it unclear how to deal with systematic and theoretical uncertainties.
 
%should we consider
%the uncertainties on the solar fluxes as statistical experimental unceratinties
%or as systematic uncertainties,
%or as theorethical uncertainties
%(that in a frequentist approach are not even allowed to exist)?
%The standard procedure consists in not raising such questions and
%summing all errors in quadrature.
%Indeed one can see that this is the right thing to do (in the Gaussian limit) 
%if one applies
%the Neyman construction in a Bayesian framework.}.

Using the analytical properties of Gaussians 
enormously simplifies the computation: we will not need lengthy and obscure computer calculations.
The probability $p$ is computed as described in the appendix.
We will study oscillations among the three active neutrinos in a two flavour setup.
We could study much more general cases, but experiments indicate that this seems to be the relevant case\footnote{
In a 3 $\nu$ framework, the $\nu_e$ can also
oscillate at the atmospheric $\Delta m^2$. The CHOOZ bound~\cite{CHOOZ} implies that the relative mixing angle is so small that it can only have a minor effect
on solar neutrinos. The LSND anomaly~\cite{LSND} motivates models with a fourth sterile neutrino.
However LSND is significantly constrained directly by Bugey~\cite{Bugey} and Karmen~\cite{Karmen}, and indirectly by SK~\cite{KaSun,SKatm} that disfavours a
significant sterile contribution to both the atmospheric and solar anomalies.
These indirect bounds can be evaded in models with many sterile neutrinos.
%Extra dimensions motivate a tower of sterile neutrinos~\cite{nuR5dA}, that give rise to active/sterile oscillations at
%an effective $\Delta m^2$ much less than the cutoff, in presence of a
%a {\em single large\/}  extra dimension.
%But we do not see how it could arise because couplings explode linearly with the radius~\cite{Antoniadis}.
%Forgetting this problem, the simplest models have serious problems with supernova bounds~\cite{amen}.
%One can still obtain an interesting phenomenology not clearly incompatible with supernova bounds
%by building contrivied models where one attempts to break in a proper way the symmetry (lepton number?)
%necessary to  avoid  a TeV-scale mass term for the ordinary neutrinos~\cite{AccanimentoTerapeutico}.
}.
$\Delta m^2 \equiv m_2^2-m_1^2>0$ is the squared mass difference relevant to solar neutrinos,
and $0\le \theta\le \pi/2$ is the corresponding mixing angle.

\begin{figure}[t]             
\begin{center}
\hspace{4cm} 90\% CL\hspace{7cm} 99\% CL\hfill~\\[2mm]
\includegraphics[width=8cm]{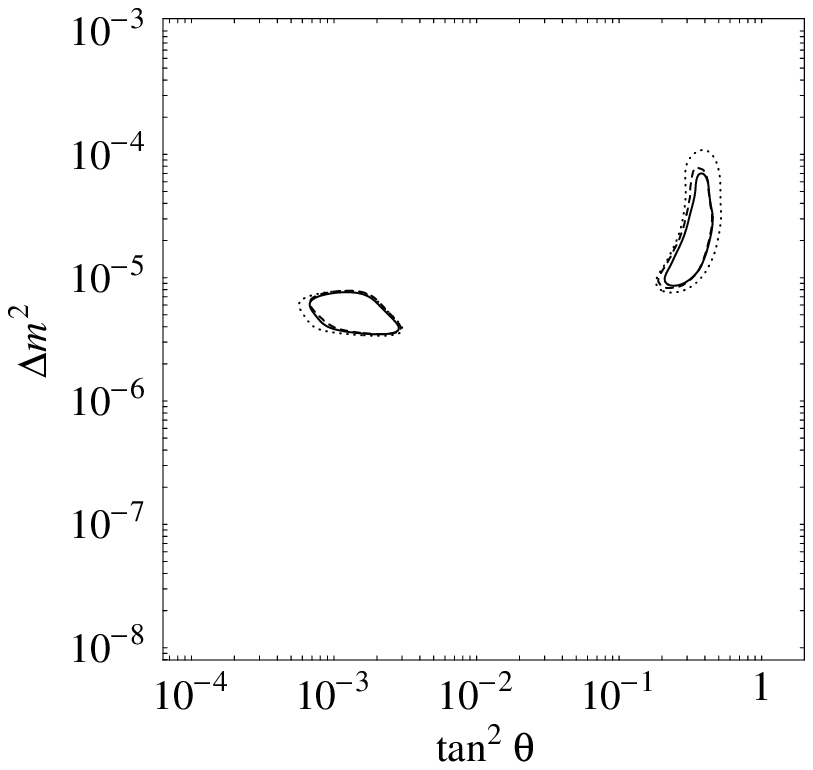} 
\includegraphics[width=8cm]{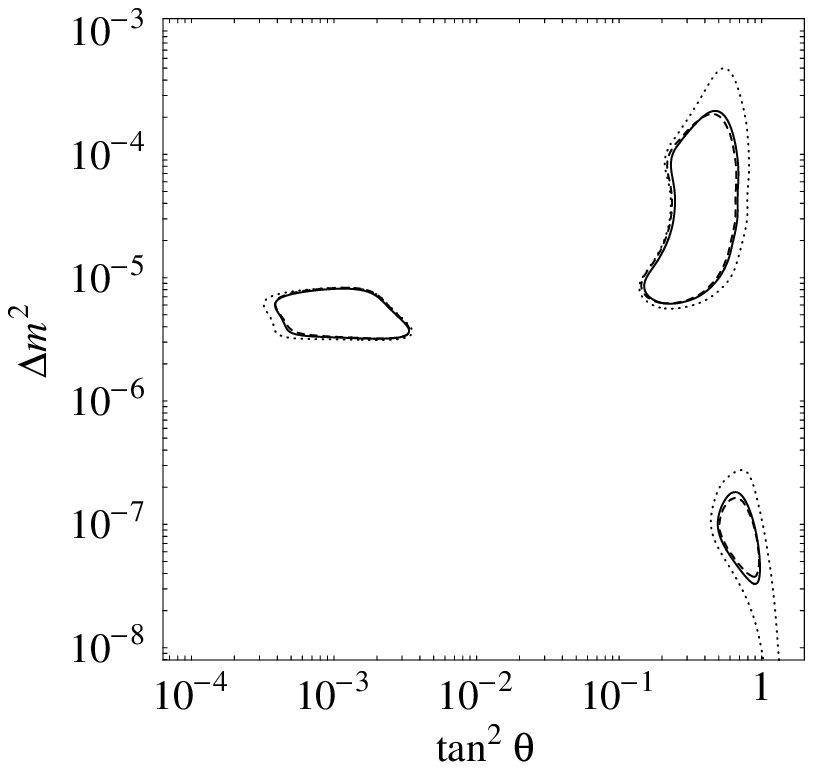}
\caption{\label{fig:piano} \em Confidence regions at 90\% (left) and 99\% (right) CL
obtained from the measured solar rates using three different methods. The smallest 
regions (continuous line) are obtained with the $\Delta\chi^2$ approximation;
they are surrounded by the Feldman--Cousins regions (dashed line). 
The largest regions (dotted line) are obtained with the Crow--Gardner procedure.}
\end{center}
\end{figure}

In section~\ref{rates} we fit the data about the total rates using the Crow--Gardner and Feldman--Cousins 
constructions, which are compared with the commonly used $\Delta \chi^2$-cut approximation. 
We do not find dramatic differences (see fig.~\ref{fig:piano}).
A fit based on the $\Delta\chi^2$ approximation does not miss any relevant physical issue.
In section~\ref{fullset} we include in the fit the SK spectral and day/night data.
We now find more marked differences between the various methods for building Neyman's confidence regions (see fig.~\ref{fig:fccg}).
In section~\ref{GOF} we show that
the well-known statement that LMA, LOW and SMA presently give a good fit
is based on an inappropriate statistical test,
and we recompute the goodness-of-fit of the various solutions (see table 1).
%of the goodness-of-fit tests for the various solutions, uncorrectly interpreted in the literature.
Our conclusions are drawn in section~\ref{conclusions}.

%  Before moving to a completely frequentist approach, we stress that the presence 
%  of different likelihood peaks in the solar neutrino analysis makes the choice of a prior distribution 
%  for the parameters rather crucial for a Bayesian study. 
%  One can choose to parametrize our ``prior ignorance'' with a flat prior pdf in $\log\Delta m^2$ and
%  $\log \theta$ variables, but $\log\tan^2\theta$ or $\log\sin^2 2\theta$ are good as well: different choices 
%  give a different relative weight to the various solutions. Therefore it is impossible, without specifying
%  (and justifying) a prior distribution, to give a quantitative sense to a Bayesian analysis.

\section{\label{rates} Different frequentist analyses: rates only}
We want to compare exact and approximate methods to compute confidence regions. To begin with, 
we consider only  the total rates measured at Homestake, 
SuperKamiokande and the weighted sum of the two Gallium experiments: GALLEX-GNO and SAGE. 
All fits done so far (except~\cite{giunti})
use the approximated method based on the $\Delta \chi^2$-cut; 
this approximation will be compared with %%%the two presently most common frequentist constructions: 
two frequentist constructions: 
the Crow--Gardner~\cite{CG} and
Feldman--Cousins~ \cite{FC} methods.

The use of the $\Delta \chi^2$-cut is based on the well-known {\em likelihood ratio theorem}~\cite{Eadie}, which states:
given a conditional pdf $p(\mb{x}|\mb{m})$ (\mb{x} is the data vector and \mb{m} are the parameters we want 
to estimate) with a range for $\mb{x}$ independent from the value of \mb{m}, the quantity
\be
\label{eq:likratio}
\lambda(\mb{x};\mb{m}) = 2 \log \left(\frac{\max_{\mb{m}} p(\mb{x}|\mb{m})}{p(\mb{x}|\mb{m})}\right) 
\ee  
is distributed as a $\chi^2$ with dim(\mb{m}) degrees of freedom (dof), independently from the value
of \mb{m}, in the limit $\dim(\mb{x})\rightarrow\infty$. With the pdf of eq.~(\ref{eq:gaussian}) this
leads to 
\be
\label{eq:lrgaussian}
\lambda(\mb{x};\dm,\theta) = \chi^2 - \chi^2_{\rm best} + \log\det \sigma^2 -\log\det \sigma^2_{\rm best}, 
\ee 
where $\chi^2$ is the usual sum in the exponent of eq.~(\ref{eq:gaussian}), $\sigma^2$ is the covariance
matrix and the subscript `best' indicates that the corresponding quantity must be evaluated at the
value of $\dm,\theta$  that maximizes the probability for the given measured $\mb{x}$. 
In the limit of infinite data,  $\lambda$ is distributed as a $\chi^2$ with two degrees of 
freedom (the two parameters we are studying: $\dm$ and $\theta$.
If
$10^{-3}\eV^2\circa{<}\Delta m^2\circa{<}10^{-4}\eV^2$ one can obtain poor fits
with energy independent survival probability.
In this case the experimental results only depend on the single parameter $\theta$). 

The simplest way
to construct confidence regions  using this asymptotic property is to include all values of $(\dm,\theta)$
for which $\lambda$ is less than a critical value, which can be obtained from the $\chi^2$ distribution tables.
Neglecting the $\ln\det \sigma^2$ term --- which is not a constant --- gives the well-known approximate rule
\be
\label{eq:chicut}
\Delta \chi^2 < \beta,
\ee
where $\beta$ depends on the confidence level (CL) we want to quote. This is the method most analyses use to obtain
the confidence regions.
It is twofold approximate: 
it neglects the $\log\det \sigma^2$ dependence and, since the number of solar data is finite, does not ensure the correct CL.
% \footnote{
% Including the $\ln\det \sigma^2$ contribution to the pdf gives small but visible corrections.
% Taking into account this ingredient we could easily perform a Bayesian analysis.
% However the arbitrariness of the prior distribution function $p(\Delta m^2,\theta)$ would give  
% an uncertainty comparable to this effect.
% Therefore, it does not seem useful to show the resulting Bayesian fit.}.

\begin{figure}[t]            
\begin{center}
\includegraphics[scale=.4,angle=0]{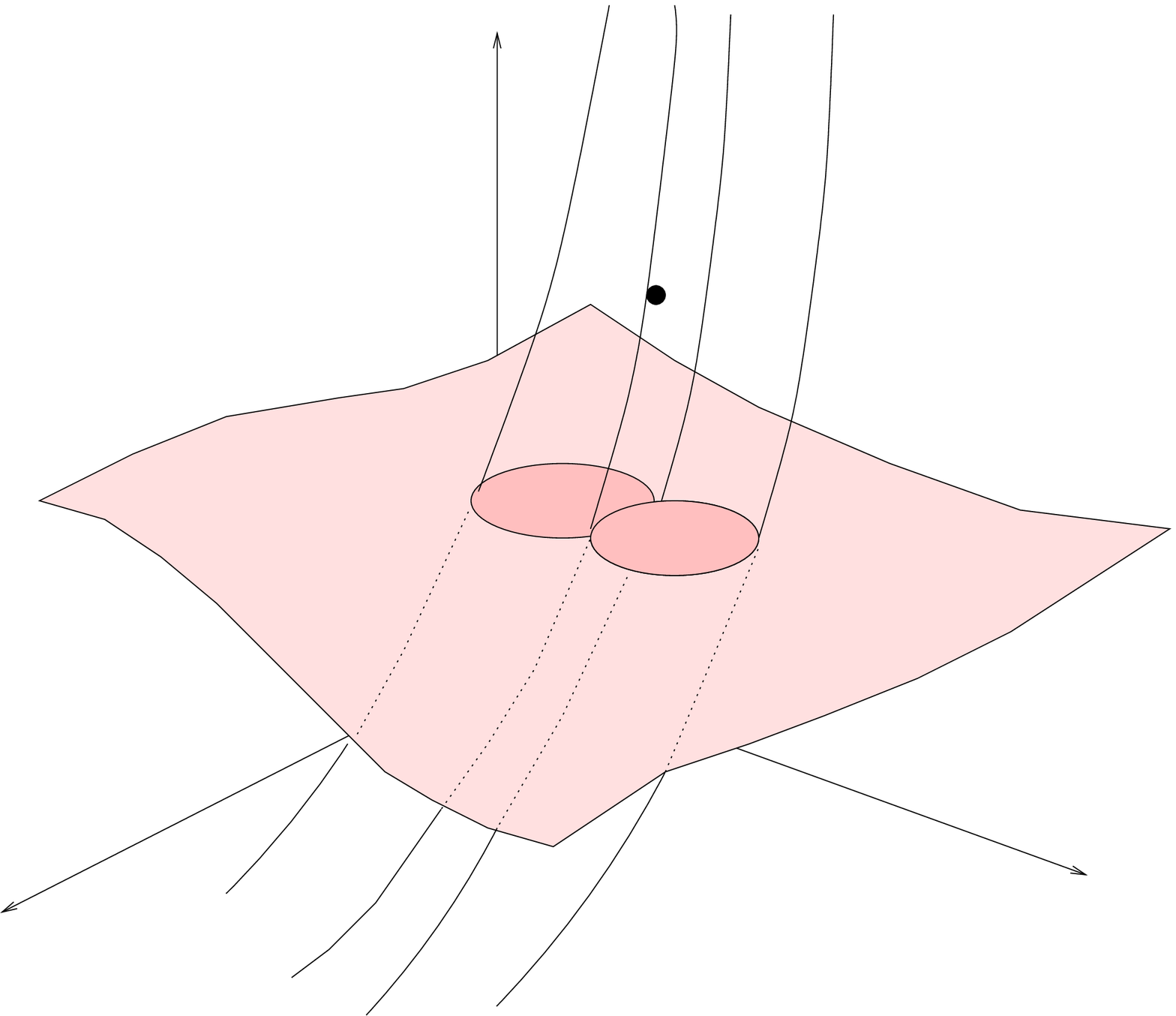}
\includegraphics[scale=.4,angle=0]{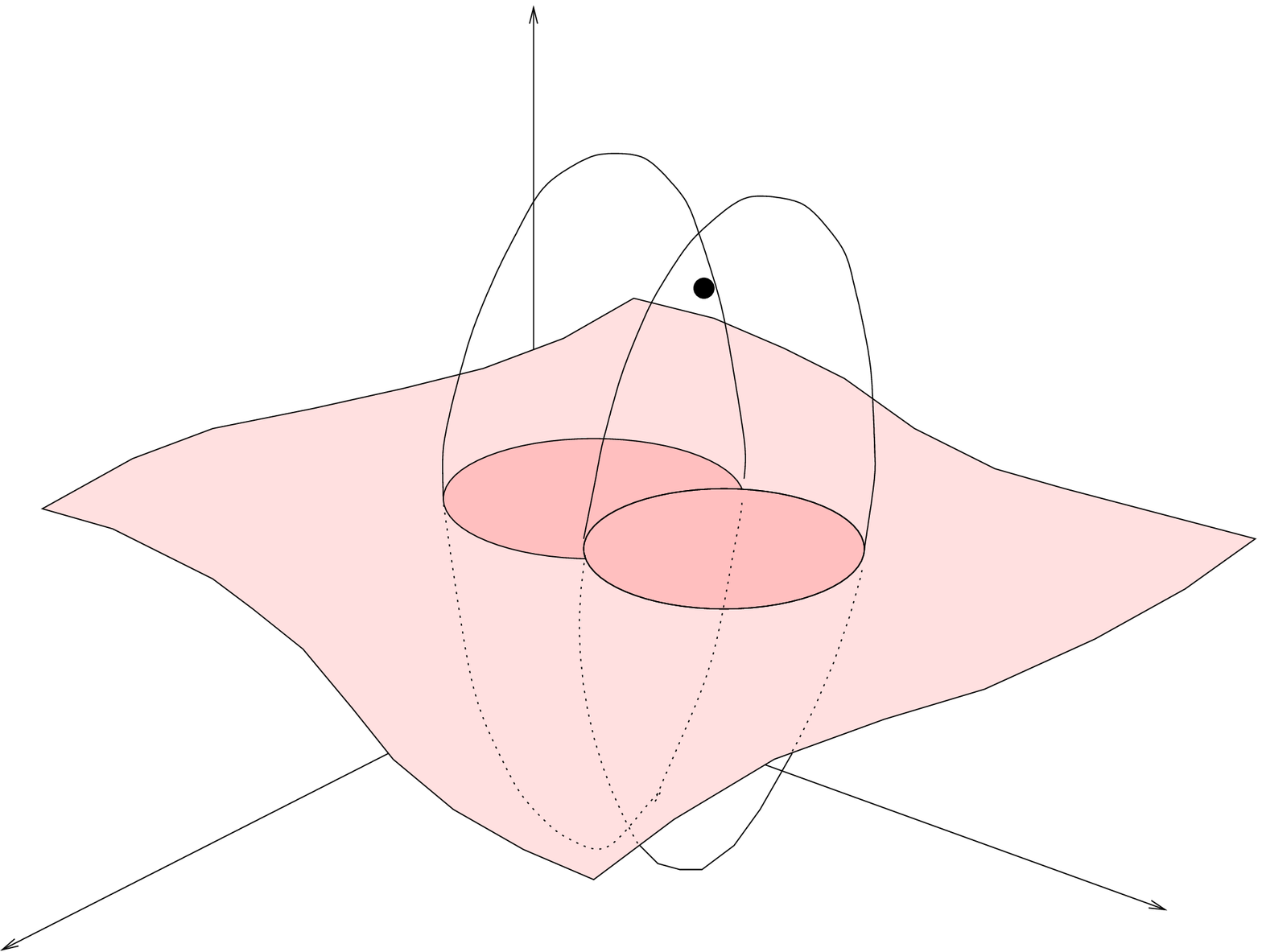}
\caption{\label{fig:cilinders} \em 
%  Approximate shapes of the acceptance regions for two near parameter points in
%  the FC case (left) and in the CG one (right). In the first case they are stretched perpendicularly to the theoretical 
%  surface, while in the second they are ellipsoidal.
%  A possible experimental point is indicated: its inclusion in the various acceptance regions defines which parameter
%  points are accepted.
Approximate shapes of the acceptance regions for two near
parameter points in the FC case (left) and in the CG case (right). In the
first case the regions are stretched perpendicularly to the theoretical
surface, while in the second they are ellipsoidal. A parameter point is
accepted if the measured experimental point (black dot in the figures)
lies inside its acceptance region.} 
\end{center}
\end{figure}

The correct frequentist construction of the confidence regions is a well-known procedure: 
for any point \mb{m} in the parameter space ($\dm$ and $\theta$, in our case) one has to {\em arbitrarily} choose a 
region ${\cal{A}}(\mb{m})$ in the space $\{\mb{x}\} $
of the data (the three rates, in our case) which contains the CL\% of the probability.
The knowledge of the pdf\eq{gaussian} allows to do this.
The confidence region ${\cal{B}}(\mb{x})$ is given by all the points in the parameter space that contains the measured value of the
experimental data in
their acceptance region:
\be
\label{eq:band}
{\cal{B}}(\mb{x}) = \{\mb{m}|\mb{x} \in {\cal{A}}(\mb{m}) \}.
\ee
It is easy to realize that, taken any {\em true} value for the parameters, the quoted region
contains this true value in the CL\% of the cases. 

\smallskip

The arbitrariness in the choice of the acceptance region ${\cal{A}}(\mb{m})$ can be fixed by choosing a particular ordering in the data space: the 
construction of ${\cal{A}}(\mb{m})$ is made adding $\mb{x}$-cells in 
that order until the requested probability (coverage) is reached. 
Crow and Gardner (CG) proposed~\cite{CG} an ordering based on $p(\mb{x}|\mb{m})$, while
Feldman and Cousins (FC) proposed \cite{FC}\footnote{The Feldman--Cousins
method supplements Neyman's prescription in a way that solves some of the problems that characterize other methods. 
Some unwanted and strange properties are
however still present, and the proposed way out~\cite{punzi} goes beyond Neyman's construction, leading, 
always in a completely frequentist approach, to the definition of a new quantity called Strong Confidence Level.} an ordering based on the likelihood ratio
\be
\label{eq:FCorder}
p(\mb{x}|\mb{m})/p(\mb{x}|\mb{m}_{\rm best}(\mb{x})).
\ee
This means that the `priority' of a point $\mb{x}$ for fixed $\mb{m}$
is given by its probability relative to the probability obtained with the parameters set at the best-fit value $\mb{m}_{\rm best}(\mb{x})$
corresponding to $\mb{x}$.

\subsection{Feldman--Cousins fit}
The FC ordering requires cumbersome numerical computations, but guarantees that the FC acceptance regions share
the nice properties of the approximate $\Delta \chi^2$-cut method.
The FC ordering disregards the statistical fluctuations with no
information on the parameters. If the measured rates are unlikely for any
value of the parameters, the FC procedure ``renormalizes'' the probability
when determining the confidence regions.     
It is easy to see that the FC acceptance regions are never empty for any choice of the confidence level:
every point in the data space  belongs
at least to the FC acceptance region of the parameter
$\mb{m}_{\rm best}$ that maximizes its probability\footnote{The Feldman--Cousins procedure gives no empty 
confidence regions if all the points with the same likelihood ratio are included in the acceptance region for 
given parameters, even after the given probability is reached: this can give a certain {\em overcovering},
but is essential to get this good property. Some pathological situations are still possible if 
the best-fit point does not exist~\cite{punzi}.}.

The FC procedure has many points in common with the
approximate $\Delta \chi^2$-cut method: looking at eq.s~(\ref{eq:likratio})--(\ref{eq:chicut}) 
at fixed \mb{m} we
see that the inequality $\lambda < \beta$ chooses ${\cal{A}}(\mb{m})$ with the same ordering
as the FC method. The only difference is that the limit $\beta$ is chosen, using the asymptotic distribution 
of $\lambda$, independent of \mb{m}, while the exact method gives a limit $\beta(\mb{m})$ that depends on the 
oscillation parameters $\mb{m}$.

It is easy to check that the $\Delta \chi^2$-cut is exactly equivalent to the FC 
construction if the pdf is Gaussian with constant covariance matrix and with theoretical rates 
that depend linearly on the
parameters (by `theoretical rates' we indicate the most likely value of the rates, for given values of $\Delta m^2$ and $\theta$).  
In the linear
approximation the theoretical rates,  obtained varying the two parameters $\dm$ and $\theta$, form a plane in the three dimensional space of the 
rates\footnote{It is interesting to note that even if a 
fundamental property of frequentist inference is its independence from the metric and topology of the parameter
space, the validity of the $\Delta \chi^2$-cut approximation strongly depends on them.}. 
The comparison with this linear approximation helps us to understand whether the $\Delta\chi^2$-cut is a good approximation. Two different behaviors are
possible for a given \mb{m}:
\begin{itemize}
\item The value of $\beta(\mb{m})$ given by the FC procedure
is smaller than the approximated one derived by the $\Delta\chi^2$-cut. This happens, for example, if we measure 
values of the parameters near the edges
of the parameter space. The $\Delta\chi^2$ approximation assumes an infinite hyperplane of theoretical rates: 
the points of the data space that have maximal probability for `non-physical' values of the parameters will be 
included in the acceptance regions of the points near the edge, reducing their limit $\beta$ to have a correct coverage. 
\item The value of $\beta(\mb{m})$  is larger than the approximated one.
This happens when different regions of the parameter space give similar predictions for the data. A data 
point that is included in the acceptance region of a given \mb{m} in the linear approximation may have a 
bigger $p(\mb{x}|\mb{m}_{\rm best})$ because of the folding of the hypersurface, which lowers its likelihood ratio. 
To reach the requested probability a bigger value for $\beta$ is needed.
\end{itemize}
Within the LMA, LOW and SMA regions the linear approximation is pretty good, as the curvature of the `theoretical
surface' is small with respect to the typical errors on the rates. 
The effects due to the edges of the theoretical surface can also be neglected. {\em The main deviation from the 
$\Delta\chi^2$ approximation is due to the fact that SMA, LMA, LOW points give similar predictions\/}: the surface of 
theoretical rates is folded.
Constructing the acceptance region for an oscillation parameter \mb{m} in the SMA region, 
we find points with best-fit parameters 
in the LMA region, so that the $\Delta\chi^2$ approximation is expected to give some undercoverage. 
The FC acceptance regions at $90\%$ and $99\%$ CL are plotted in figure \ref{fig:piano}: we see that the regions obtained from the $\Delta \chi^2$-cut are
smaller than the exact FC  regions. 
For example, the approximated $\Delta \chi^2$ cut at 90\% CL is $\Delta \chi^2< 4.6$.
The value of $\beta(\mb{m})$ at 90\% CL obtained with the FC construction is 
$\approx (4.6 \div 5.5)$ for $\mb{m}$ in the SMA region and $\approx(4.8 \div 5.5)$ for $\mb{m}$ in the LMA region.
Furthermore, owing to the variation of $\log\det \sigma^2$ (neglected by the $\Delta\chi^2$ approximation),
in the SMA and LOW regions the FC boundary
intersects the $\Delta\chi^2$ boundary, instead of surrounding it.
The difference between the approximate and rigorous methods is however small enough to  justify the $\Delta \chi^2$ approximation.

\subsection{Crow--Gardner fit}
%Plottare la superficie teorica??
A second way to construct confidence regions is based on the CG ordering: the acceptance regions
are built beginning from the points of highest probability.
Such ordering is not invariant under a reparametrization of the experimental data
(e.g.\ a CG fit of the rates is different from a CG fit of the squared rates).
Such acceptance regions are the smallest with
the given coverage. 
The difference between the FC and the CG procedures is essential for
those points that are unlikely for any value of the parameters, i.e.\ all the points far from the surface of
theoretical rates.
We have seen that, with the FC ordering, 
every data point is included in the acceptance region of at least one point in the parameter space, but this is 
obviously not true for the CG  method. Consider for example the linear approximation in which the theoretical rates describe 
a plane in the rate space.
For a given \mb{m}, in view of~(\ref{eq:gaussian}), 
the CG acceptance region will be an ellipsoid centered in the most likely value for the rates, 
which becomes larger and larger with growing CL. The FC acceptance regions will be very different and stretched to 
infinity in a cylindrical shape perpendicularly to the plane (see figure \ref{fig:cilinders}). 
This is clear if we consider that in this approximation
the maximum likelihood point $\mb{m}_{\rm best}$ is obtained by projecting a data point on the 
`theoretical plane' (in the base where the covariance matrix is
proportional to the identity): all
the points lying on a line perpendicular to this plane and intersecting it in the point described by \mb{m} have 
likelihood ratio equal to 1 and are included in the acceptance region.

\begin{figure}[t]            
\begin{center}
{\includegraphics[bb=0 0 558 500,scale=.4,angle=0]{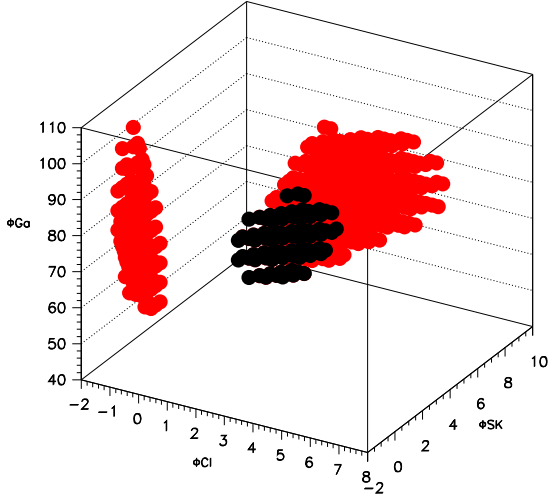}}
\hspace{1cm}
\raisebox{1.5cm}{\parbox[b]{5cm}{\caption{\label{fig:banda} \em Comparison between the CG acceptance region (black) and the FC one (red) 
at 90\% CL for a parameter point 
in the SMA region. Chlorine and Gallium rates are in SNU, while the SK rate is in $10^6 \,{\rm cm}^{-2} {\rm s}^{-1}$.
The presence of LMA points with comparable predictions makes the FC region asymmetric and disconnected.}}}
\end{center}
\end{figure}

In figure \ref{fig:banda} 
we compare the FC and CG acceptance regions for one given SMA oscillation. We
see that the CG region is ellipsoidal as expected, while the FC one is stretched, but in one direction only. This is
due to the strong non-linearity of the `theoretical surface': LMA and SMA have similar rate predictions even if they have
very different parameters. The surface is folded and this causes the asymmetry in the FC region: the acceptance region
is deformed and disconnected to get far from the LMA predictions.

The CG acceptance regions at $90\%$ and $99\%$ CL are plotted in figure \ref{fig:piano}.
The differences between the FC and CG confidence regions are readily understood. If we fix the experimental data and begin from
a very low CL, we expect an empty CG region (there is no ellipsoid that contains the data) while the FC region
is small but non empty. As the CL increases, a CG region appears (roughly at 4\% CL in our case) and all the regions grow.
With a large CL we expect the CG regions to be larger than the FC ones as the ellipsoids have a larger projection
on the `theoretical surface' than the stretched FC acceptance regions (see fig.~\ref{fig:cilinders}). 
All these features can be checked explicitly in figure \ref{fig:piano}. 

In conclusion two points must be stressed: first of all we have checked that a correct frequentist approach gives results 
only slightly different from the naive analysis based on the $\Delta \chi^2$-cut. 
This is apparently in contrast to what is obtained in~\cite{giunti}. The main difference between that analysis and
ours is that we use all three rates to construct confidence regions, while \cite{giunti} finds, with a Monte Carlo
simulation, the distribution of the maximum likelihood estimators $\Delta\hat{m}^2$ and $\hat\theta$ and construct from
this the confidence regions, using the Crow--Gardner ordering. Since  the two $\Delta\hat{m}^2$, $\hat\theta$
are not a {\em sufficient statistics} for the three rates, this procedure implies a certain loss of information, which 
leads to larger confidence regions.

% Comparison with Giunti: computable loss of information? 

A second point is the comparison between the two methods, CG and FC: the results are pretty 
similar. As we will see in the next section, this is no longer the case when the SK data on the angular and energy 
distribution are included in the fit.  
We have not shown `vacuum oscillation' fits of the solar rates
because they are strongly disfavoured by the SK data.

\section{\label{fullset} The inclusion of the whole data set}
The SuperKamiokande collaboration has also measured the
energy spectrum of the recoil electrons as a function of the zenith-angle position of the sun.
The full data set usually employed in solar neutrino fits contains 38 independent dof (see the appendix).
With such a number of data it is practically impossible to perform a complete numerical construction of the acceptance regions,
without any approximation. For example, even if we divide every dimension of the data space in only 20
cells, we arrive to a 38-dimensional space divided into $\sim 10^{50}$
cells. For this reason we cannot construct the FC confidence regions with all the data set. However, 
for the same reasons described in the previous section,
the approximated  $\Delta\chi^2$-cut method is expected to be a reasonable approximation of the FC construction and
to give confidence regions slightly smaller than the FC ones.

\begin{figure}[t]            
\begin{center}
\includegraphics[width=55mm]{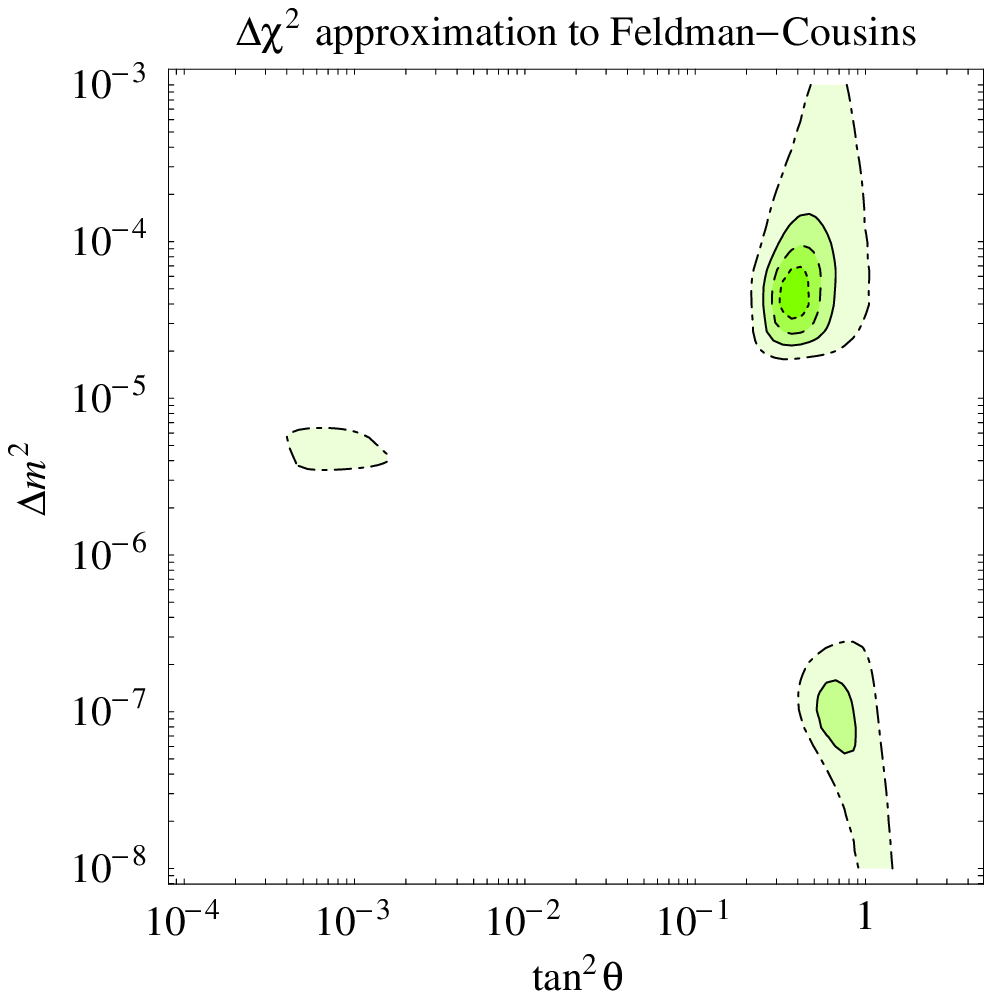}
\includegraphics[width=55mm]{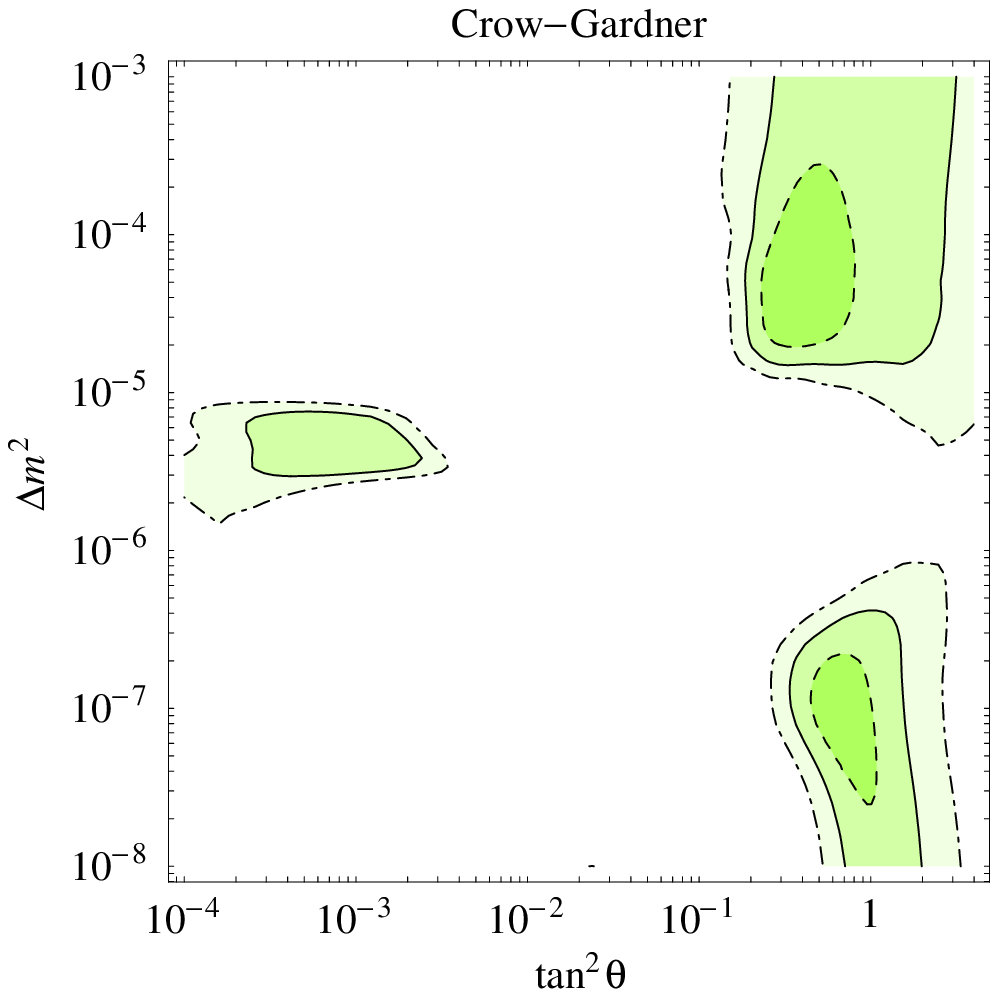}
\includegraphics[width=55mm]{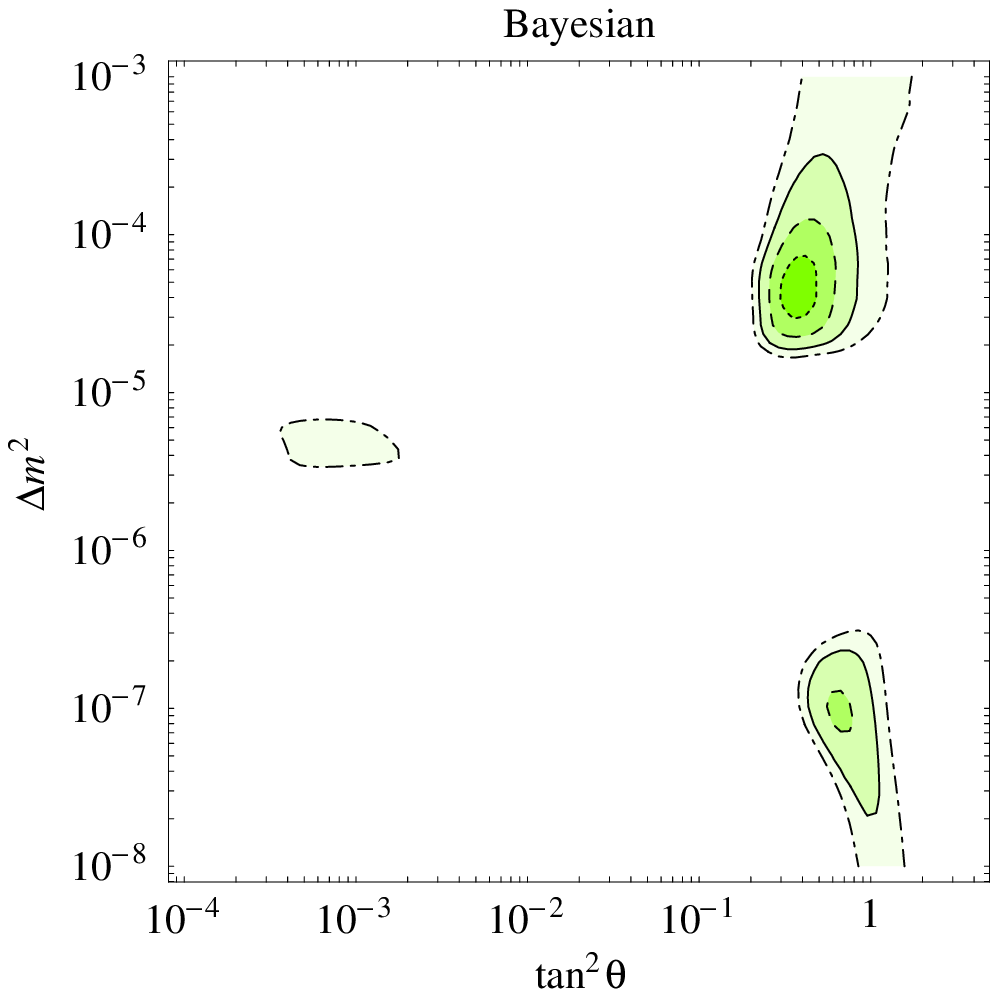}
\caption{\label{fig:fccg} \em Comparison between the $\Delta \chi^2$-cut (left), the CG method  (middle)
and a Bayesian fit done assuming a flat prior in the plotted variables (right). We show the
35\% (dotted lines), 68\% (dashed), 90\% (continuous) and 99\% CL (dot-dashed) regions. }
%The CG 35\% region is absent but the CG regions at high CL are much greater than the corresponding FC regions.}
\end{center}
\end{figure}

For the CG ordering the situation is better. Since we have approximated the pdf~(\ref{eq:gaussian}) as a Gaussian function of the data, the CG
construction is equivalent to a cut on the $\chi^2$ with 38 dof (rather than on the $\Delta\chi^2$). 
Note that in this case the procedure is exact even if $\log\det \sigma^2$ is not constant.
For any \mb{m} the $\chi^2$-cut defines an
% log det \sigma^2 con 25 dof: grandi cambiamenti???
ellipsoidal acceptance region ${\cal{A}(\mb{m})}$,
and the confidence region is given by the set of parameters with $\chi^2$ smaller 
than a given value. The comparison between the CG ordering and the approximated $\Delta\chi^2$-cut can be done
analytically: for any given value of CL
\begin{equation}\begin{array}{lrl}
{\rm FC}~\approx \Delta\chi^2{\rm-cut}:\qquad & \chi^2(\dm;\theta) -\chi^2_{\rm best} & \le {\rm Quantile}(\chi^2_{\rm \:2\:dof},{\rm CL}) \\
{\rm CG}~= \chi^2{\rm-cut}: & \chi^2(\dm;\theta) & \le{\rm Quantile}(\chi^2_{\rm \:38\:dof},{\rm CL}).
\end{array}
\end{equation}
The comparison between the CG method and the $\Delta\chi^2$-cut, which we can consider an approximation of the FC
ordering, is shown in figure \ref{fig:fccg} 
and presents all the features described in the previous section. 
The differences are now rather evident.
The CG regions are empty until the $\approx$ 40\% CL, while the \dchi\ regions are never empty (as the FC ones).
% The CG regions, depending on the $\chi^2$ distribution with 38 dof, grow much faster than the others: in the case considered .
The two methods give equal regions
for $\sim$ 45\% CL. With a larger CL the CG regions are bigger than the $\Delta\chi^2$ regions.
Figure~\ref{fig:fccg} shows that the arbitrariness in constructing frequentist acceptance regions can be quite significant. 
For example a CG fit accepts a large $\dm \approx 10^{-3} \eV^2$ at $\approx 75\%$ CL, while 
the $\Delta\chi^2$ approximation to the FC fit rules it out at $ \approx 98\%$ CL. 
One should keep in mind the statistical assumptions behind this limit, when using it
to demonstrate the necessity of a hierarchy between the squared-mass differences characteristic of the solar and
atmospheric anomalies.
We now explain the reason of this difference and argue that the FC fit is the relevant one.

It could seem strange that, after including all data, the CG confidence regions with high CL are much larger than when fitting only the rates.
This happens because the CG method does not use the information on the parameters contained in the data in the most efficient way.
Statistical  fluctuation leads to experimental results that do not lie on the theoretical surface.
The CG ordering, treating
in the same way the directions in the data space `perpendicular'' and `parallel' to the surface,
does not use the information contained in the `distance' from the theoretical surface to obtain further information on the parameters.
This is why the CG ordering leads to strange results, though correct from the coverage point of view.
The FC method uses the `distance' from the theoretical surface to recognize and
eliminate the statistical fluctuations that have nothing to do with the determination of the parameters.
This difference between the two procedures is much more significant when fitting the full data set than when fitting the rates only.
There are now many more data than unknown parameters: the theoretical surface is two-dimensional in a 38-dimensional data space.

%%A CG fit employs the information concerning the position of the experimental point {\em along} the theoretical surface, 
%%while the ``distance'' from the surface is not important. 

In figure \ref{fig:fccg} we also show a Bayesian fit, done assuming a flat prior $dp(\Delta m^2,\theta)=d\ln\Delta m^2\,d\ln\tan\theta$.
Unlike the CG fit, this Bayesian fit should coincide with the $\Delta\chi^2$ approximation to the FC fit if the pdf were a 
Gaussian function of $\ln\Delta m^2$ and $\ln\tan\theta$. 
By comparing the two fits we can again see visible but not crucial corrections due to the non-Gaussianity in 
the theoretical parameters $\Delta m,\theta$.
The arbitrariness in the prior distribution function $p(\Delta m^2,\theta)$ gives
an uncertainty comparable to the effect due to the non-Gaussianity.

\section{\label{GOF} The goodness of fit}
A goodness-of-fit (GOF) test studies the probability of the experimental data $p({\rm data}|{\rm theory})$ under given
theoretical hypotheses  (a model of the sun, the assumption that neutrinos are oscillating rather than decaying\ldots) 
leading to statements of the form: if all the hypotheses are true, 
the probability that the discrepancy between predictions and data is due to statistical fluctuations is less than a certain amount.
The purpose is to understand if the theoretical hypotheses
used to explain the data are plausible or not.

\begin{table}[t]
$$
\begin{array}{c|ccc}
\parbox[c]{2cm}{\centering \Blue Goodness\\ of fit\Black} & 
\hbox{\Blue ($a$) Rates only} &
\parbox[c]{4cm}{($b$) Rates and spectra:\\\phantom{(b) }naive result}&
\parbox[c]{4cm}{($c$) Rates and spectra:\\ \phantom{(c) }refined result}\Black \\ \hline
\rm\Blue  SMA\Black &  55\% ( 58\%\cite{recentfits1},  50\%\cite{recentfits2},51.8\%\cite{giunti}) & 30\% ~(34\%\cite{recentfits2}) &  \circa{<}~2\,\% \\
\rm\Blue  LMA\Black &  ~6\% ( 10\%\cite{recentfits1},  ~8\%\cite{recentfits2}, 6.3\%\cite{giunti}) & 60\% ~(59\%\cite{recentfits2}) &  \circa{<}15\,\% \\
\rm\Blue  LOW\Black & 0.7\% (1.6\%\cite{recentfits1}, 0.5\%\cite{recentfits2}, 1.1\%\cite{giunti}) & 50\% ~(40\%\cite{recentfits2}) &  \circa{<}~2\,\% \\
\rm\Blue  \hbox{$P_{ee}={\rm cte}$}\Black & 0.3\%& 28\%  & \circa{<}~1\,\% 
% entrano nella best fit region a 3%, 100%, 15% e 1% C.L.
\end{array}$$
\caption{\em GOF of the SMA, LMA, LOW and energy-independent solutions obtained by fitting (a) only the rates,
(b) including all the data, (c) including only the `most significant' data.
The energy-independent solution contains only one free parameter, while SMA, LMA, LOW have two free parameters.
Our results are compared with the corresponding ones in~\cite{recentfits1,recentfits2,giunti}.
The symbol $\circa{<}$ recalls that the GOF values could be slightly lower.}
\end{table}

When analyzing the most recent data, one encounters the following paradoxical situation:
the LOW solution gives a {\em poor\/} fit of the solar rates only 
(e.g.\ \cite{recentfits2} finds a GOF of $0.5\%$).
After including the full data set, LOW gives a {\em good} fit (e.g.\ \cite{recentfits2} finds a GOF of $40\%$).
The paradox is that we have added 35 zenith and energy bins, in which there is no signal for neutrino oscillations.
It is clearly necessary to understand better the meaning of the GOF test before we can decide if
LOW gives a decent fit or not.
This important question also applies to the other alternative solutions.
Such tests are based on Pearson's $\chi^2$: the quantity~\cite{Eadie}
\be
\sum_{ij} (x_i^{\rm exp} - \hat{x}_i^{\rm th}) \hat{\sigma}^{-2}_{ij} (x_j^{\rm exp} - \hat{x}_j^{\rm th})
\ee   
is  asymptotically 
% Perche' Parapa dice esattamente?
distributed as a $\chi^2$ with $N_{\rm data} - N_{\rm param}$ dof\footnote{The result is exact if the variation of $\log\det \sigma^2$ can be neglected.
Furthermore, with a finite number of data, this test is exact only with theoretical rates depending linearly on
the parameters.
The deviation from this approximation leads to a small overestimate of the GOF:
for example, fitting only the rates, the GOF of SMA gets corrected from $51.8\%$ to $48.4\%$~\cite{giunti}.}.
The hats indicate that the corresponding quantity must be evaluated at the maximum likelihood parameter point.
The CG ordering is deeply linked to this GOF test: the absence of a confidence region
until a given CL value is correlated to the goodness of the fit.

The paradoxical increase of the GOF of LOW is
clearly due to the fact that the Pearson test does not recognize that
there is a problem concentrated in the three solar rates that contain all the evidence for neutrino oscillations.
It only sees that the total $\chi^2$ is roughly equal to the large total number of dof (38), so that the fit seems good.
According to this procedure, even
a global three neutrino fit of solar, atmospheric and LSND data would seem good.
It is well known that the LSND anomaly requires a fourth neutrino, or something else.

We now explain why the Pearson's test is not adequate for such a situation.
Pearson's test does a precise thing: it tests the validity of a certain
solution with respect to a generic alternative hypothesis, which has a sufficient number of parameters to fit all the
data with infinite precision. 
Therefore the inclusion of more data changes also
the set of alternative hypotheses which we compare with.
{\em Describing the recoil electron spectrum in terms of 18 energy bins implies that
we admit alternative theories with fuzzy energy spectra\/}.
No physical mechanism could generate an irregular spectrum, so that we do not want to test this aspect.
The measured spectrum is of course regular, and Pearson test rewards the LOW solution for this reason.
To better understand this point,
suppose that we add as new data the direction of arrival of the interacting neutrinos.
All solutions (including the no-oscillation hypothesis) would have a higher GOF.
It is obvious that these solutions are much better than a generic one, because they at least `know' where the Sun is.
A meaningful $\chi^2$ test should include only those data that really test the hypothesis under consideration.
On the contrary, the inclusion of irrelevant data does not affect the confidence regions built with the
FC ordering, 
so that a naive application of the $\Delta\chi^2$-cut correctly approximates the best-fit regions.

We therefore conclude that testing the goodness of the fit using a lot of energy bins
gives a formally correct answer to an irrelevant question.
If a smaller set of data were used to describe the spectral and angular information, the set of alternative hypotheses would 
be more reasonable and one would conclude that there is a goodness-of-fit problem.
%  The fact that SK uses energy bins about 3
%  times smaller than the energy resolution of the detector is not the crucial point.
%  More importantly,
Most of the information on the energy and zenith-angle spectra
can be condensed into observables such as the mean recoil electron energy and the day/night asymmetry,
as shown in fits presented by the SK collaboration~\cite{ExpsSun}.

Within our assumption of 2-neutrino oscillations,
the main new information encoded in SK spectral and day/night data
is that the survival probability $P_{ee}(E_\nu) $ can only have a mild energy dependence 
around $E_\nu\sim 10\MeV$.
This can be seen in a simple way by parameterizing  $P_{ee}(E_\nu) $ as
\begin{equation}\label{eq:P'0}
P_{ee}(E_\nu) = P_0  +  P'_0 \cdot \bigg(\frac{E_\nu}{10\MeV} -1\bigg) + \frac{P''_0}{2}\bigg(\frac{E_\nu}{10\MeV} -1\bigg)
^{\!\!2}+\cdots
\end{equation}
The SK spectral data measure $P'_0 = -0.05\pm 0.1$ and disfavour
the SMA solution because it prefers a larger slope $P'_0$.
Significant non-linearities in $P_{ee}(E_\nu)$ are not predicted in SMA, LMA, LOW oscillations, nor could be recognized by SK
(the  present error on $P''_0$ is $\sim 1$).

%(in SNO the measurable recoil electron energy $T_e$ will instead be more strongly correlated with $E_\nu$).
In conclusion most of the information contained in the SK spectral data can be conveniently
condensed into a single observable $f$, that measures the slope of
$P_{ee}(E_\nu)$. One possible choice is
the ratio between the rate of `low energy' (i.e. $T_e < 9\MeV$) and `high energy' (i.e.\ $9\MeV<T_e<13\MeV$) recoil electrons,
as measured by SK.
The upper bound on the recoil electron energy $T_e$ has been chosen in order to avoid potential problems due to an enhanced flux of h$e$p neutrinos.
The measured value and the uncertainty on $f(\rho_i)$ can be easily deduced from the SK data on the full energy spectrum:
$\sigma^2_f = f_i \sigma_{ij}^2 f_j$ (in Gaussian approximation), 
where $f_i \equiv \partial f/\partial \rho_i$,  $\{\rho_i\}$ is the full set of SK bins
and $\sigma^2_{ij}$ is the full error matrix.

By supplementing the fit of the total neutrino rates with a single observable $f$ we find the GOF values shown in table~1.
The symbol $\circa{<}$ recalls that, especially in the SMA case, 
it could be possible to obtain slightly lower GOF values by identifying
another observable more sensitive to the energy dependence of the neutrino survival probability.
However, a variation of the GOF between, say, $1\%$ and $4\%$, is within the
uncertainty due to arbitrariness inherent in any statistical analysis.
The important point is that the GOF values are significantly lower than the values based on a naive $\chi^2$ test,
and motivate a non-standard analysis of the solar neutrino anomaly~\cite{noSSM}.
The SK collaboration~\cite{KaSun} finds that SMA now gives a poor fit using another reasonable procedure:
at 97\% CL, the region favoured by total rates falls inside the region disfavoured by spectral and day/night data.

\section{\label{conclusions} Conclusions}
Fits of solar neutrino data are usually done using the $\Delta\chi^2$-cut valid in the Gaussian approximation and find
few distinct best-fit solutions (LMA, SMA, LOW, \ldots).
Since a Gaussian would have only one peak, it is useful to check the validity of
the $\Delta\chi^2$-cut approximation by comparing its results with the
exact confidence regions built using the Neyman construction.
This has been done using two different ordering prescriptions, proposed by Crow--Gardner (CG) and by Feldman--Cousins (FC).
{\em We find that the $\Delta\chi^2$ cut provides a good approximation to the Feldman--Cousins confidence regions.\/}

When the full data set is used, there is some significant difference between the CG and FC regions.
Even if the two  methods give regions with the same statistical meaning, their conceptual  significance is different. 
The FC regions are not influenced by statistical fluctuations with no information on the oscillation parameters,
while the CG regions are composed by all the oscillation parameters that provide an acceptable fit of the data.
The meaning of the results is deeply influenced by the assumptions involved in the statistical analysis.
We finally show that a correct understanding of the meaning of Pearson `goodness-of-fit' test invalidates the
statement that all solutions (LMA, LOW and SMA) presently give a good fit.

We hope that our refined statistical analysis has been useful for clarifying some aspects of solar neutrino fits.
Only refined experimental data will allow to identify the solution of the solar neutrino problem.
We plan to update the electronic version of this paper, if new significant experimental data will soon be presented.

\begin{figure}[t]             
\begin{center}
\hspace{4cm} 90\% CL\hspace{7cm} 99\% CL\hfill~\\[2mm]
 \includegraphics[width=7cm]{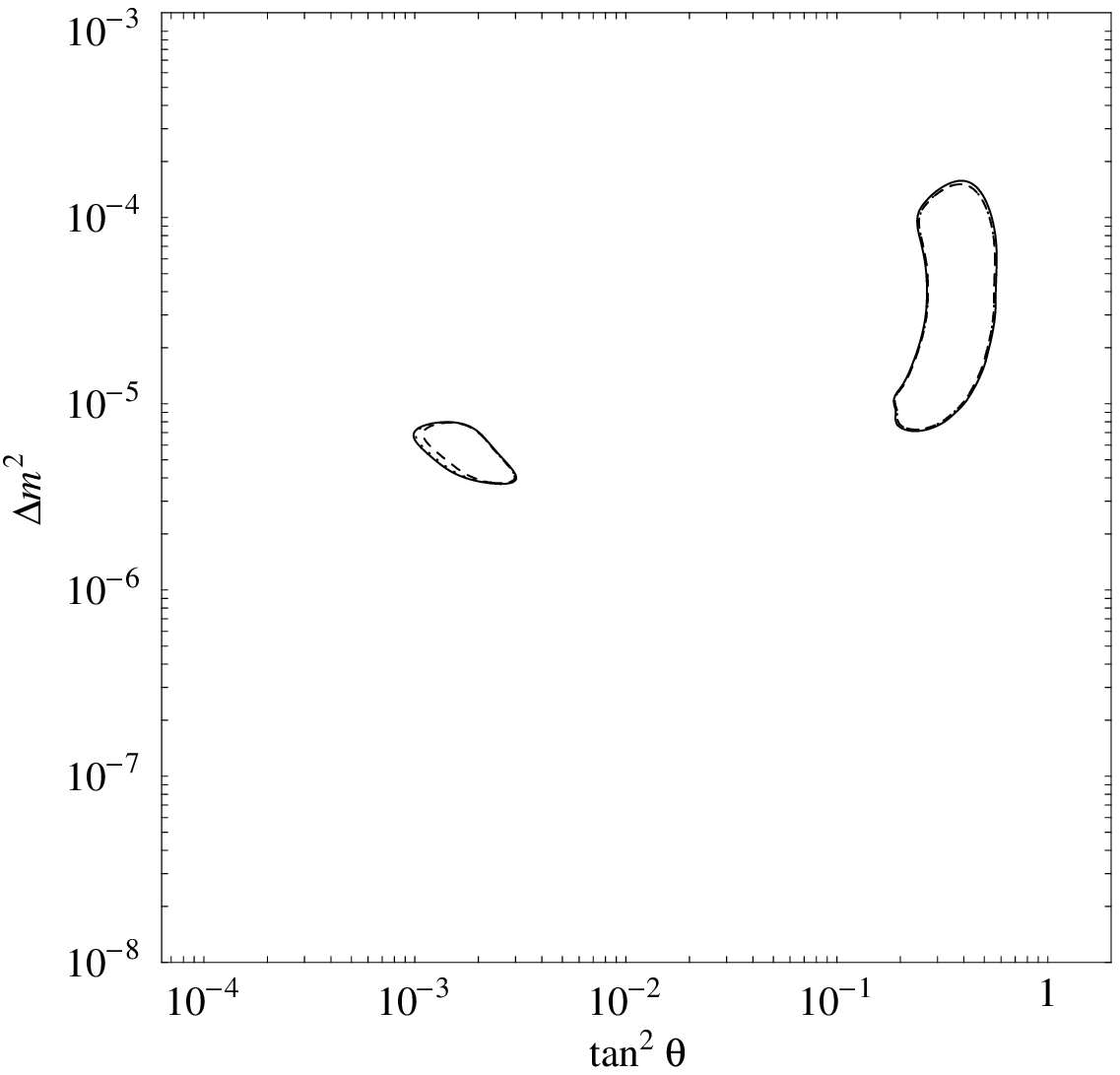} \hspace{0.8cm}
 \includegraphics[width=7cm]{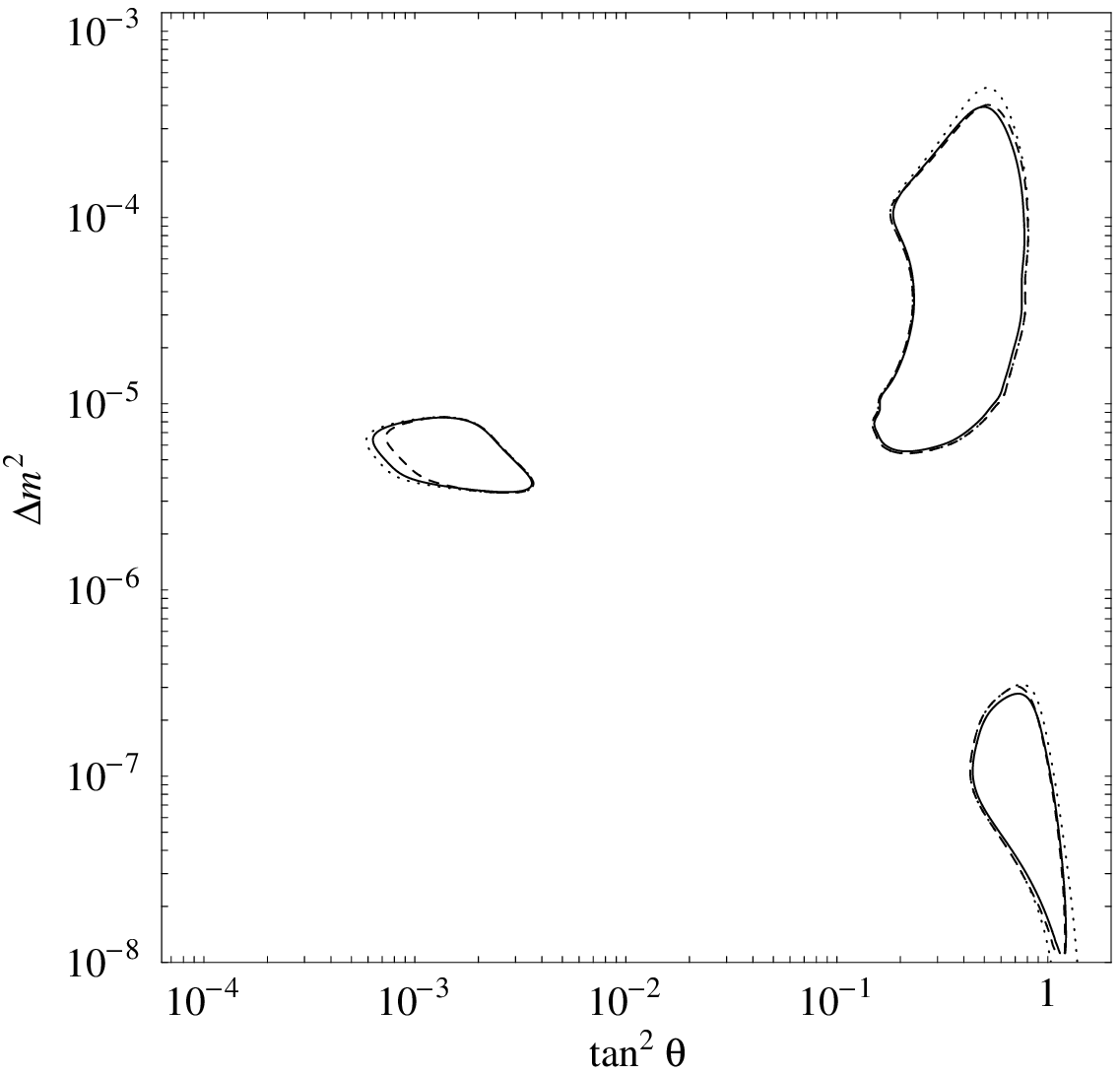}
\caption{\label{fig:ratesSNO} \em The `rates only' fit of fig.~\ref{fig:piano}, updated including the SNO results.
Confidence regions at 90\% (left) and 99\% (right) CL
obtained from the four solar rates using three different methods:
the $\Delta\chi^2$ approximation (continuous line),
the Feldman--Cousins procedure (dashed line) and
the Crow--Gardner procedure (dotted line).}
\end{center}
\end{figure}

\begin{figure}[t]            
\begin{center}
\includegraphics[width=17.5cm,height=7cm]{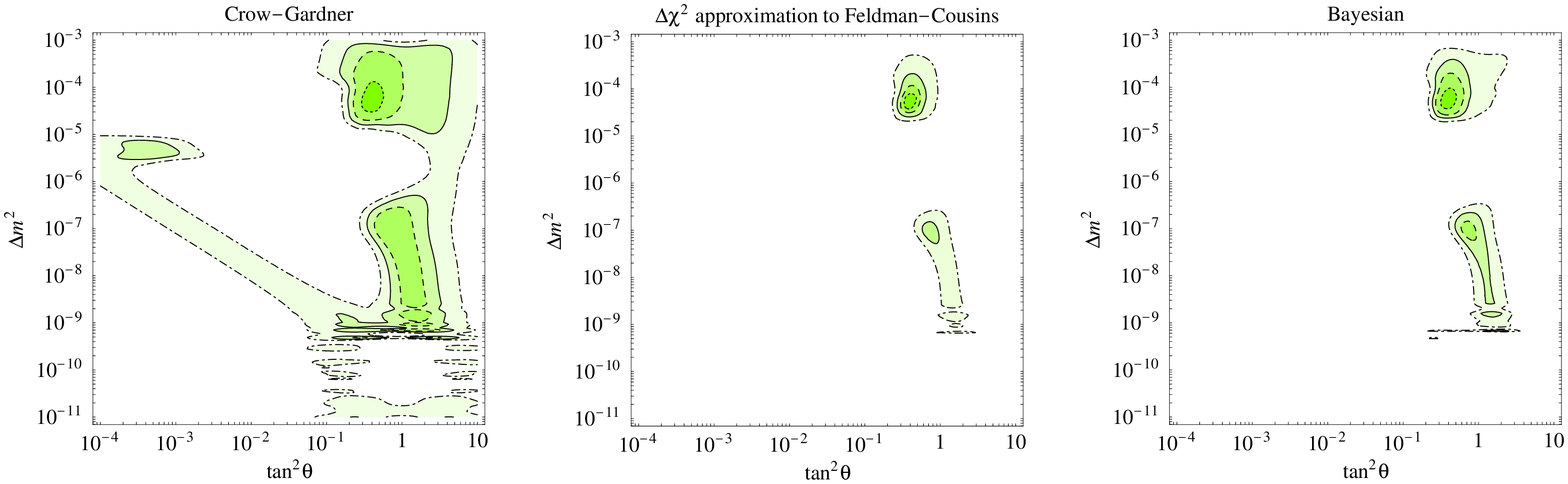}
\caption{\label{fig:globalSNO} \em The `global' fit of fig.~\ref{fig:fccg}, updated including the SNO results.
Comparison between the the CG method (left), the $\Delta \chi^2$-cut (middle),
and a Bayesian fit done assuming a flat prior in the plotted variables (right). We show the
35\% (dotted lines), 68\% (dashed), 90\% (continuous) and 99\% CL (dot-dashed) regions.}
\end{center}
\end{figure}

\section{Addendum: CC SNO results}\label{6in}
In this addendum, we  update our results by adding:

\begin{itemize}
\item the $\nu_e {\rm d} \to {\rm pp}e$ (CC) rate measured at SNO~\cite{SNO} above the energy threshold 
$T_{\rm eff}> 6.75\MeV$: $R_{\rm SNO}^{\rm CC} =
(1.75\pm 0.13) 10^6\,{\rm cm}^{-2}{\rm s}^{-1}$.

\item the latest  SK results~\cite{SKnew} (1258 days). 
The total flux now is $R_{\rm SK} = (2.32\pm0.08)\cdot 10^{6} \,{\rm cm}^{-2}{\rm s}^{-1}$.
The day/night energy spectrum contains a new bin with energy $5\MeV< E_e < 5.5\MeV$.

\item Furthermore,
we explicitly include the CHOOZ data~\cite{CHOOZ} in the global fit.
CHOOZ data just disfavour values of $\Delta m^2$ above $0.7~10^{-3}\eV^2$ for large mixing $\theta\sim \pi/4$
and are irrelevant at smaller $\Delta m^2,\theta$.

\end{itemize}
The updated fit of the four Chlorine, Gallium, SK and SNO rates (without including spectral data)
is shown in fig.~\ref{fig:ratesSNO}.
The Feldman-Cousins SMA region is somewhat
smaller than what obtained from the $\Delta\chi^2$ approximation: this is mainly due to the
variation of the determinant of the covariance matrix in the SMA
region (see section 2).
Anyway there is no significant discrepancy between the frequentist fits and the $\Delta\chi^2$
Gaussian approximation.

Fig.~\ref{fig:globalSNO} shows the results for the global fit, including all the rates and the day/night spectrum
measured by SK.
The Crow-Gardner (CG) confidence regions are composed by all oscillation parameters with a high goodness-of-fit,
evaluated in a na\"{\i}ve way.
As explained in section 3 the CG procedure
gives very large confidence regions because it is an inefficient (although statistically correct) procedure.
The Feldman-Cousins (FC) procedure gives smaller regions because it is more efficient: unlike the CG procedure it
recognizes and eliminates the statistical fluctuations with no information on the oscillation parameters.
In the Gaussian approximation, it should agree with the Bayesian fit with flat prior distribution:
the differences between the $\Delta\chi^2$-approximation and the
Bayesian  are visible in fig.~\ref{fig:globalSNO} but not significant.

How strongly is now the SMA solution disfavoured?
Remembering that the difference in the best-fit ${\chi}^2$ is a powerful statistical indicator,
the main results can be summarized as
\begin{equation}\label{eq:postSNO}
\hat{\chi}^2({\rm SMA}) = \hat{\chi}^2({\rm LMA}) + \left\{\begin{array}{ll}
14.0 & \hbox{our fit}\cr
14.7 & \cite{LisiSNO}\cr
10.1 & \cite{BPGSNO}\cr
11.2 & \cite{indiaSNO}
\end{array}\right.
,\qquad
\hat{\chi}^2({\rm LOW}) = \hat{\chi}^2({\rm LMA}) + \left\{\begin{array}{ll}
3.2 & \hbox{our fit}\cr
4.5 & \cite{LisiSNO}\cr
3.1 & \cite{BPGSNO}\cr
4.4 & \cite{indiaSNO}
\end{array}\right.
\end{equation}
where, for comparison, we also report the results of three other post-SNO analyses.
%  Rate dependent uncertainties are computed using the thorethical values, rather than using the experimental values.
Like in~\cite{BPGSNO}
 and unlike in~\cite{LisiSNO,indiaSNO} we improved the definition
of the $\chi^2$ by using as data the SK flux in the $18+18$ energy day/night bins,
rather than separating the SK data into total flux and un-normalized spectrum.
This improved $\chi^2$ takes into account statistical correlations between the total SNO and SK rates and the SK spectral data.
Due to the different energy thresholds in SK and SNO such kind of correlations are now
non negligible\footnote{A different effect that shifts the SMA region by a comparable amount is the variation of the $\det\sigma^2$ term
in eq.s\eq{gaussian},\eq{lrgaussian}.}:
using the improved $\chi^2$ the difference $\Delta\hat{\chi}^2 \equiv \hat{\chi}^2({\rm SMA}) - \hat{\chi}^2({\rm LMA})$ 
is reduced\footnote{The SMA solution is still present at $99\%$ CL only in the plots in~\cite{BPGSNO} due to
a simpler reason: they used the $\Delta\chi^2$ corresponding to 3 degrees of freedom
(rather than to 2 degrees of freedom as here and in~\cite{LisiSNO,indiaSNO})
because their analysis was made in a model with a fourth sterile neutrino and `$2+2$' mass spectrum,
where solar oscillations depend on one more parameter.} by  $\sim 3$.
%  As in~\cite{BPGSNO} the total SK flux obtained by summing the flux in each bin
%  is 1 standard deviation lower than the value quoted by SK.
%   In the gaussian limit the $\Delta\hat{\chi}^2$ in\eq{eq:postSNO}
%  would be distributed as a $\chi^2$ with zero degrees of freedom:
A MonteCarlo simulation would be necessary to determine precisely how $\Delta\hat{\chi}^2$
is distributed: presumably\eq{postSNO} implies that the evidence for LMA against SMA is now somewhere between
 ``$3\sigma$'' and ``$4\sigma$''.

As discussed in section~4, the claims in~\cite{BPGSNO,indiaSNO} that the SMA solution
(and few other disfavoured solutions) still give a `good fit'
is based on an inefficient goodness-of-fit test   % unable of recognizing a bad fit.
on the value of some global $\chi^2$.
In the case of parameter determination,
the Crow-Gardner procedure is statistically correct
(like the global $\chi^2$ test),
but no one employs it just because it is inefficient
(as clear in fig.~\ref{fig:globalSNO}).
Using the more efficient goodness-of-fit test discussed in section~4, we get
the updated post-SNO results
$$\hbox{GOF(LMA)}\sim 25\%,\qquad
\hbox{GOF(LOW)}\sim 4\%,\qquad
\hbox{GOF(EI)}\sim 1\%,\qquad
\hbox{GOF(SMA)}\sim 0.4\%,$$
where EI denotes the `energy independent' solution obtained for $\Delta m^2 \circa{>} 2\cdot 10^{-4}\eV^2$.
Unlike few years ago LMA, LOW or EI oscillations are now better than SMA 
not because new data gave new significant positive indications for them,
but because new data (compatible with no spectral distortion)
gave a strong evidence against SMA.
The `worse' solutions of few years ago are now the `better' ones,
but cannot give very good fits.
Of course all results in this paper are based on the (sometimes questionable) assumption
that all systematic and theoretical uncertainties involved in the game have
been correctly estimated.
The final results do not significantly change if
the (not yet calibrated) Chlorine result is not included in the fit.

\begin{floatingfigure}[r]{8cm}
$$\begin{picture}(7,6.5)
\put(0,0){\includegraphics[width=7cm]{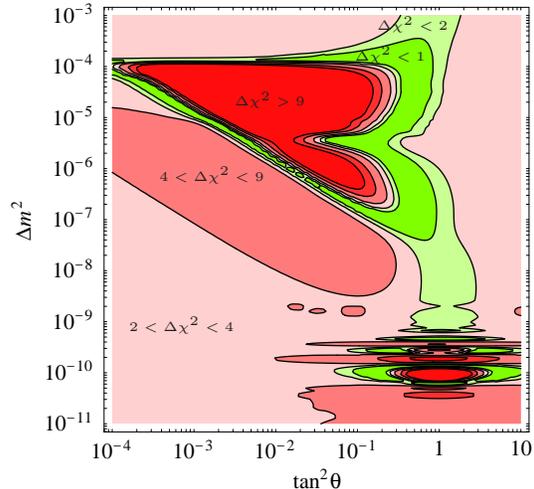}}
\put(4.6,6){\tiny$\Delta \chi^2 < 1$}
\put(4.9,6.4){\tiny$\Delta \chi^2 < 2$}
\put(2,4.4){\tiny$4<\Delta \chi^2 < 9$}
\put(1.6,2.4){\tiny$2<\Delta \chi^2 < 4$}
\put(3,5.4){\tiny$\Delta \chi^2 > 9$}
\end{picture}$$\vspace{-1cm}
\caption{\label{fig:deltaSNO} \em Regions `favoured' (green) and `disfavoured' (red)
by the first SNO data.}\vspace{0.5cm}
\end{floatingfigure}
In order to show in more detail the impact of SNO, 
in fig.~\ref{fig:deltaSNO} we plot the values of the variation
in the $\chi^2$ due to the SNO data:
$$\Delta\chi^2 = \chi^2(\hbox{post-SNO})- \chi^2(\hbox{pre-SNO}).$$
We see again that SNO results perfectly agree with LMA and LOW
(for the value of the Boron flux predicted by solar models)
and disfavour the SMA solution.
Energy-independent solutions are also slightly disfavoured
with respect to LMA and LOW:
neglecting possible small corrections due to $\theta_{13}$,
energy-independent oscillations
can at most convert half of the initial $\nu_e$ into $\nu_{\mu},\nu_\tau$ 
(for maximal mixing),
while SNO and SK indicate that the $\nu_{\mu},\nu_\tau$ flux is $\sim 1.5$ standard deviation
higher than the $\nu_e$ flux.

Finally, it is useful to recall another important result:
performing a global analysis including all experimental data (solar, atmospheric, reactor, short-baseline)
in  models with a fourth sterile neutrino,
the best $\chi^2$ is obtained
when the sterile neutrino is {\em not\/} used at all
(so that atmospheric and solar data are explained by oscillations into active neutrinos,
while the LSND data are not explained).
This means that
the present LSND evidence~\cite{LSND} for effects due to a sterile neutrino is weaker than the present evidence
of other experiments against it.
\label{6out}

\section{Addendum: NC and day/night SNO results}\label{7in}
\begin{figure}[t]             
\begin{center}
\includegraphics[width=8cm]{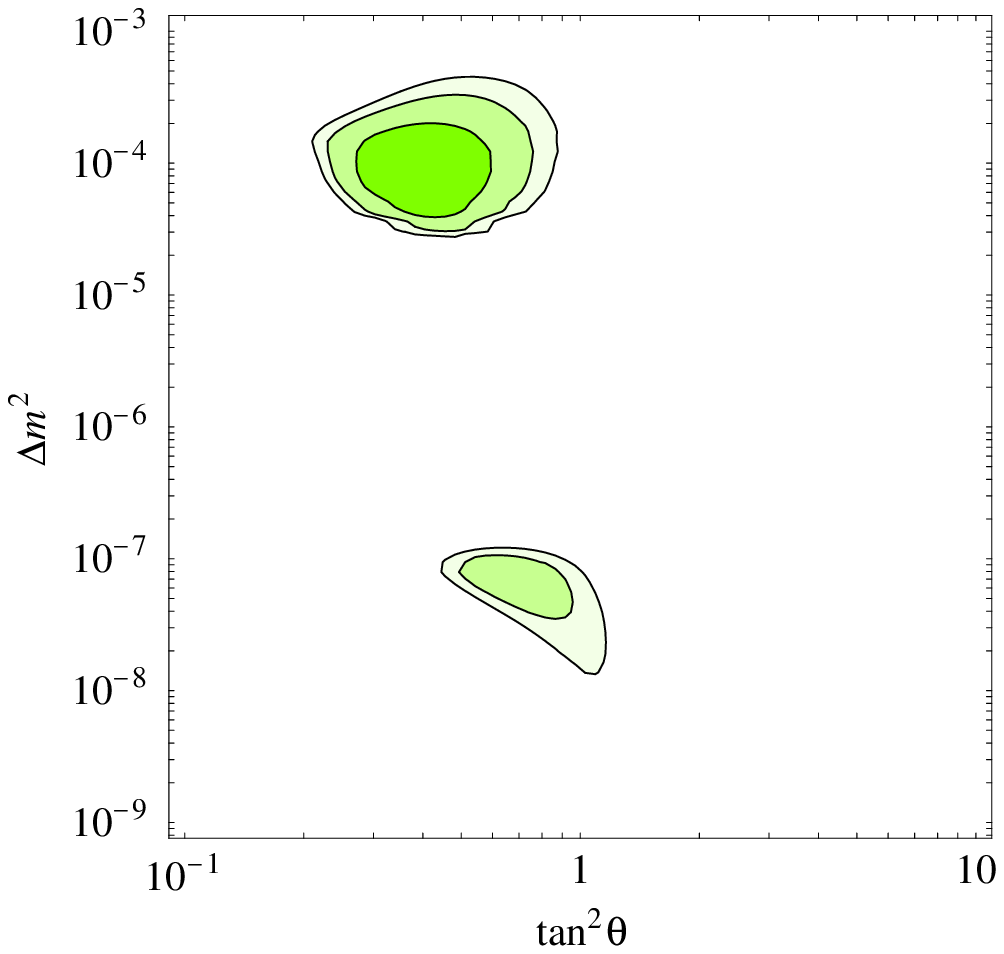} 
\includegraphics[width=8cm]{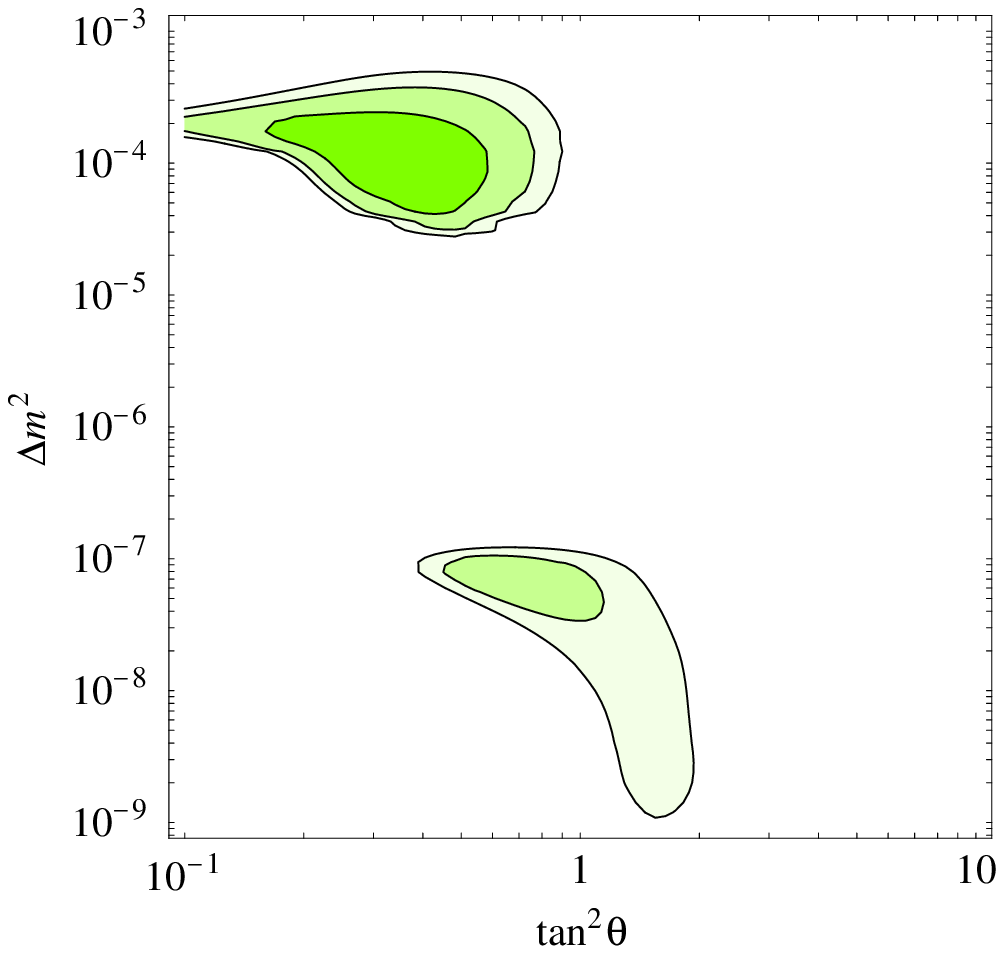} 
\caption[X]{\label{fig:globalSNO2} \em 
Best fit regions at $90$, $99$, $99.73\%$ {\rm CL}
assuming $\nu_e\leftrightarrow \nu_{\mu,\tau}$ oscillations.
In fig.~\ref{fig:globalSNO2}a
we fit the
Cl, Ga rates, the SK day/night and zenith angle spectra,
the SNO NC and CC rates and day/night asymmetries and
the CHOOZ energy spectrum,
assuming the BP00~\cite{BP98} solar model.
In fig.~\ref{fig:globalSNO2}b we drop the
BP00 prediction for the Boron and Beryllium solar fluxes.}
\end{center}
\end{figure}

In this addendum, we  update our results by adding:

\begin{itemize}

\item The $\nu {\rm d} \to \nu{\rm pn}$ (NC) and the $\nu_e {\rm d} \to {\rm pp}e$ (CC) rate
measured at SNO~\cite{SNONC}
during the initial pure D$_2$O phase using events with measured \v{C}erenkov energy, $T_{\rm eff}$, larger than $5\MeV$.
{\em Assuming} an energy-independent survival probability, $0\le P_{ee}\le 1$,
% no distorsion in the $^8$B energy neutrino spectrum, 
SNO claims
\begin{eqnsystem}{sys:SNO}
R_{\rm SNO}^{\rm CC} &=&
1.76\pm 0.06\hbox{(stat)}\pm0.09\hbox{(syst}) ~10^6\,{\rm cm}^{-2}{\rm s}^{-1} = \Phi_{^8{\rm B}} P_{ee},\\
R_{\rm SNO}^{\rm NC} &=&
5.09\pm 0.44\hbox{(stat)}\pm0.46\hbox{(syst})~10^6\,{\rm cm}^{-2}{\rm s}^{-1}= \Phi_{^8{\rm B}} (1-\eta_{\rm s}(1 - P_{ee}))
\end{eqnsystem}
(with some mild anti-correlation;
$0\le \eta_{\rm s}\le 1$ is the eventual sterile fraction involved in solar oscillations)
giving a
a $5.1\sigma$ direct evidence for neutrino conversion,
$R_{\rm SNO}^{\rm CC}/R_{\rm SNO}^{\rm NC}  < 1$.
If there are no sterile neutrinos ($\eta_{\rm s} = 0$)
$R_{\rm SNO}^{\rm NC} = \Phi_{^8{\rm B}}$, confirming
the solar model prediction for the Boron flux.
Without assuming that $P_{ee}(E_\nu)$ is energy-independent, the error
considerably increases~\cite{SNONC}
$$R_{\rm SNO}^{\rm NC} = 6.4\pm 1.6\hbox{(stat)}\pm0.6\hbox{(syst})~10^6\,{\rm cm}^{-2}{\rm s}^{-1}.$$
In order to make an oscillation fit one needs to extract $R^{\rm NC}_{\rm SNO}$ from the SNO data,
taking into account the energy distortion predicted by oscillations.
We do this using the data reported in fig.\ 2c of~\cite{SNONC}\footnote{We
here give more details about the procedure we follow.
Since SK tells that a significant spectral distortion is not present,
in order to obtain a precise fit of reasonably allowed solutions
we only need to compute the {\em shift} in the central value of $R^{\rm NC}_{\rm SNO}$.
Since the NC signal are $\gamma$ rays of energy $6.25\MeV$ (detected through their 
\v{C}erenkov photons),
events with $T_{\rm eff}> 8\MeV$ are dominantly given by CC scattering
(plus a small ES component, precisely measured by SK,
and also measured by SNO looking at forward scattering events, fig.\ 2a of~\cite{SNONC}).
The NC rate is then roughly obtained by subtracting the predicted CC rate from the number of events
with $T_{\rm eff}<8\MeV$.
Since most (about $2/3$) of these events are due to CC scattering,
%  one easily understands the errors in eq.~(\ref{sys:SNO}), and notices that 
a small spectral distortion in the CC spectrum
can give a sizable shift of the NC rate.
Including this effect the $\chi^2$ for a typical LMA (LOW) (SMA) oscillation increases by $1/2$ (0) (3)
with respect to a fit na\"{\i}vely done using eq.s~(\ref{sys:SNO}).
In presence of significant sterile effects, our approximate procedure 
could be not  accurate enough if the NC efficiency at SNO strongly depends on $E_\nu$.
We use the NC and CC cross sections from~\cite{Kubodera}.}.

% (CC:NC 2:1 explains the systematic anticorrel).
% (the CC rate can also be extracted by looking at the direction of the recoil electron).

\item The day/night asymmetries measured at SNO.
Assuming oscillations of only active neutrinos, SNO extracts
the $\nu_e$ asymmetry to be $A_e = 7.0\%\pm 5.1\%$~\cite{SNOdn}.
At $E_\nu\sim 10\MeV$ earth matter effects
are too large for $\Delta m^2$ values between LMA and LOW
(that are therefore excluded)
and are still present in the lower $\Delta m^2$ range of LMA
(here the regeneration effect is larger at high energy and for neutrinos
that cross a sizable part of the earth)
and in the upper $\Delta m^2$ range of LOW
(here the regeneration effect is larger at low energy
and has a characteristic oscillating zenith angle behavior)~\cite{recentdn}.
Both SK and SNO see a $\sim 1.5\sigma$ hint for a regeneration effect,
but neither SK nor SNO data show a statistically significant energy or zenith-angle dependence
of the effect,
that would allow to discriminate LMA from LOW.

\item The latest  SK results~\cite{Smy} (1496 days): the full 
zenith angle and energy distributions have now been released,
so that SK data now disfavour more strongly oscillations parameters that give rise to
detectable earth matter effects at SK (the lower part of the LMA solution
and the upper part of the LOW solution).
We use as SK data the rates in the 44 bins,
whose content and uncertainties are precisely
described in tables~1 and~2 of~\cite{Smy}.

\item The latest SAGE result~\cite{SAGEnew}.
The averaged Gallex/SAGE/GNO rate is now 
$R_{\rm Ga} = (72.4 \pm 4.7)\,{\rm SNU}$
(in place of the older value in eq.\eq{gasig}),
making the LOW solution less disfavoured.

\item According to solar models, the total flux of $^8{\rm B}$ neutrinos
is proportional to $S_{17}(0)$
(that parameterizes the $^7{\rm Be}\,{\rm p}\to {}^8{}{\rm B}\, \gamma$
cross section at zero energy).
New recent measurements
$$S_{17}(0) = 17.8\pm 1.3\,\rm{eV~b}~\cite{Davids},\qquad
S_{17}(0)=22.3\pm 0.9\,\rm{eV~b}~\cite{Jung},\qquad
S_{17}(0) = 22.7\pm 1.2\,\rm{eV~b}~\cite{ISOLDE}
$$
do not fully agree. Therefore we still use 
$S_{17}(0)= 19^{+4}_{-2}\,\rm{eV~b}$, as in BP00~\cite{BP98}.
After the SNO NC measurement,
this arbitrary choice is relevant almost only in fits with sterile neutrinos.

\end{itemize}
The result of our global fit is shown in fig.~\ref{fig:globalSNO2}a
(capabilities of SNO have been discussed in~\cite{preSNO}).
The best fit lies in the LMA region
$$\hat{\chi}^2\hbox{(47~dof)} = 33.0\qquad \hbox{at}\qquad
\Delta m^2 = 10^{-4.1}\eV^2,\qquad
\tan^2\theta = +0.35.$$
The LOW solution %  predicts an average $P_{ee}$ above the SNO central value and therefore
begins to be significantly disfavoured,
because the precise SNO determination of the $^8{\rm B}$ flux makes more statistically evident
that LOW oscillations give a poor fit of the other rates\footnote{The SAGE/Gallex/GNO rate
is higher than what predicted by LOW oscillations.
It will be interesting to see the next GNO data.}:
$$\hat{\chi}^2_{\rm LOW} = \hat{\chi}^2_{\rm LMA} +8.1$$
(a difference of $10.7$ is found in~\cite{SNOdn}, where
the full SK zenith-angle/energy spectra are not employed).
If LMA is the true solution,
KamLand could confirm it within few months.

\begin{figure}[t]             
\begin{center}
\includegraphics[width=8cm]{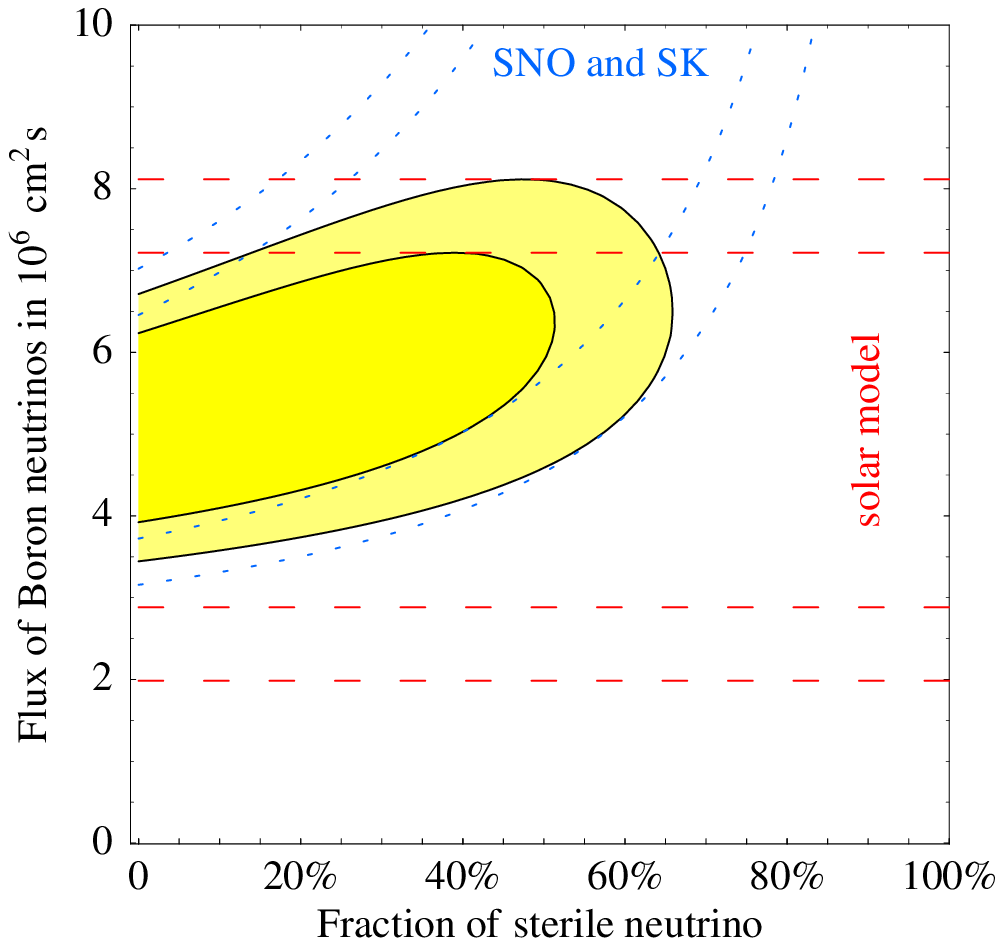} 
\includegraphics[width=7.7cm]{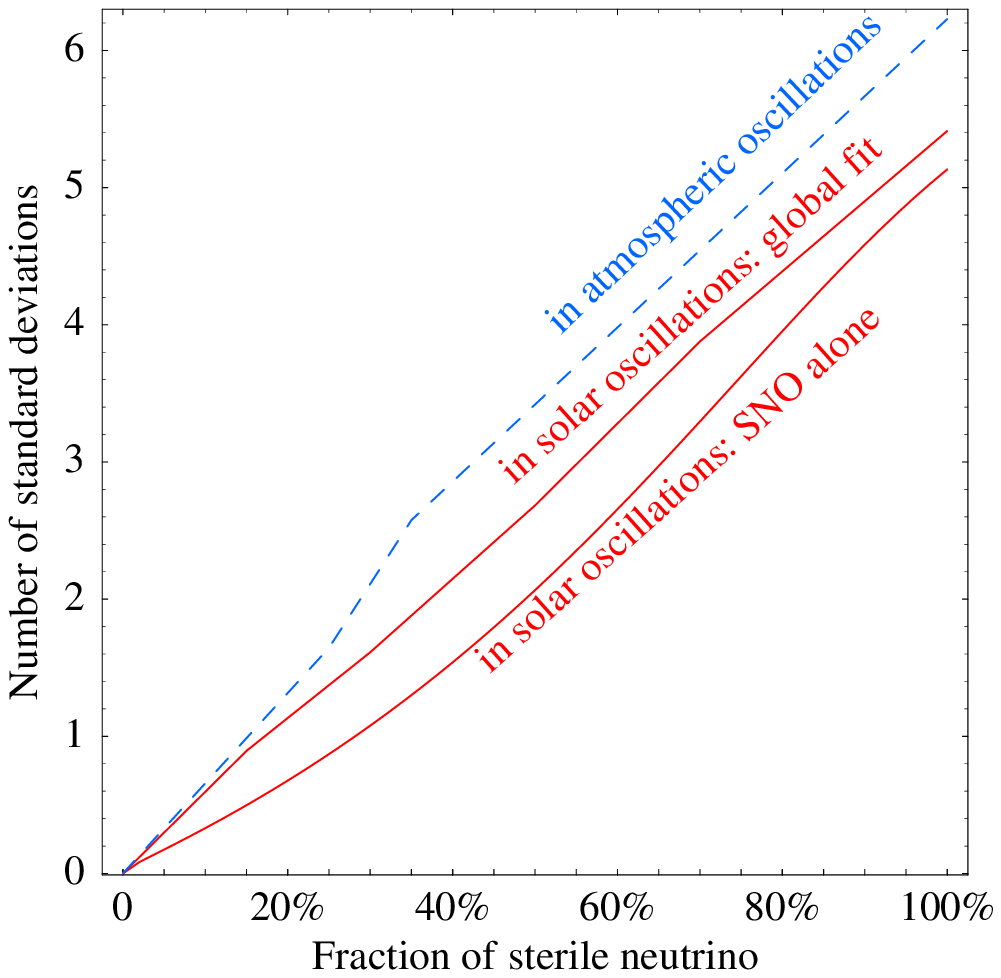}
\caption[X]{\label{fig:sterile} \em 
Fig.~\ref{fig:sterile}a: best-fit regions of the Boron flux and of the fraction $\eta_{\rm s}$
of sterile neutrino
involved in solar oscillations
at $90$ and $99\%$ {\rm CL} (2 dof).
Fig.~\ref{fig:sterile}b: fraction of sterile neutrino (1 dof) allowed in solar (continuous lines) and atmospheric (dashed line)
oscillations: in both cases data prefer no sterile neutrino.
}
\end{center}
\end{figure}

Unlike in older fits, the upper part of the LMA region
(that in the two-neutrino context we are considering gives rise to energy-independent oscillations with $P_{ee}\ge 1/2$)
is now disfavoured also by solar data alone
(the inclusion of the CHOOZ constraint now only has little effect),
since SNO suggests $P_{ee}<1/2$.
This can be considered as the first hint for MSW effects~\cite{MSW}.
% In fact, performing a more general fit
% where the solar $\nu_e$ also have 
% a mixing $\theta_{13}$ and $\theta_{\rm s}$
% with heavy ($\Delta m^2 > 10^{-3}\eV^2$)
% $\nu_{\mu,\tau} and sterile neutrinos
% one obtains a global
% $\chi^2(\Delta m^2_{12},\theta_{12},\theta_{13},\theta_{\rm s})$
% with 4 dof.
% When marginalized with respect to 
% $\theta_{13}$ and $\theta_{\rm s}$
% one obtains essentially the same $\chi^2(\Delta m^2_{12},\theta_{12})$
% (2 dof) plotted in fig.~\ref{fig:globalSNO2}.

Vacuum oscillations predicted spectral distortions not seen by SK.
Before SNO, the remaining marginally allowed
vacuum oscillation solution
was `JustSo$^2$' (with $\Delta m^2\circa{<} 10^{-11}\eV^2$)
that predicted $P_{ee}\approx 1$ (i.e.\ approximately no oscillations)
at energies probed by SK and SNO.
Oscillations with $\Delta m^2\circa{<}10^{-9}\eV^2$ 
now give the relatively better vacuum oscillation fit;
their $\chi^2$ is larger than in the best LMA fit
by about 14.

In fig.~\ref{fig:globalSNO2}b we show the results of a fit where
we do not include the BP00~\cite{BP98} predictions for the Boron {\em and} Beryllium fluxes:
the best-fit regions do not significantly change.
From the point of view of oscillation analyses,
the Beryllium flux is now the most relevant, not fully safe, solar model prediction
(see~\cite{noSSM} for a discussion):
according to BP00~\cite{BP98}
the percentage error on the Beryllium flux is only two times smaller than the one on the Boron flux.
Since Beryllium and Boron neutrinos are generated by
different scattering processes of $^7{\rm Be}$,
the agreement of the Boron flux with SNO data
suggests (but does not imply) a Beryllium flux also in agreement with solar models.

We now explain in simple terms which data 
eliminate the other solutions allowed in older fits.

\subsubsection*{Autopsy of no oscillations}
The NC and CC data from SNO alone exclude this case at $5.1\sigma$.
The statistical significance of the solar neutrino anomaly increases
to $8\sigma$, according to
our global fit.

\subsubsection*{Autopsy of small mixing angle oscillations}
For small mixing angle and in the SK and SNO energy range, $E_\nu\sim (5\div 15)\MeV$,
the $\nu_e$ survival probability is well approximated by
the Landau-Zener level-crossing probability,
$P_{ee} \approx e^{-\bar{E}/E_\nu}$,
where $\bar{E}$ is a constant (fixed by
the oscillation parameters and by the solar density gradient).
In order to get $P_{ee}\sim 1/2$ as demanded by SNO and SK
one needs to choose $\bar{E}\sim 10\MeV$.
This implies $dP_{ee}/d\ln E_\nu \sim 1$,
in contrast with the undistorted energy spectrum
observed by SK (see eq.\eq{P'0}) and SNO.
Performing a global fit we find
$$\hat{\chi}^2({\rm SMA}) = \hat{\chi}^2({\rm LMA}) +23$$
i.e.\ a $\circa{>} 4\sigma$ evidence for LMA versus SMA.
As explained in section~4, the fact that $\chi^2/\hbox{dof} \approx 1.2$ for the SMA
solution does not mean that SMA oscillations give a good fit:
the Pearson's $\chi^2$ test is not efficient 
in recognizing bad fits when there are many dof.
A more efficient test (see sections~4 and 6) confirms that SMA has a low goodness-of-fit,
as expected.

\subsubsection*{Autopsy of pure sterile oscillations}
Assuming no distortion of the energy spectrum of Boron neutrinos, SNO sees a $5.1\sigma$ evidence
for $\nu_{\mu,\tau}$ appearance.
This evidence decreases down to $3\sigma$ allowing for
generic spectral distortions, but significant spectral distortions
are not allowed by SK.
The sterile SMA solution (that gave the best sterile fit,
before SNO) gives a spectral distortion
that increases the SNO evidence for $\nu_{\mu,\tau}$ appearance,
and is therefore excluded in two different ways.
In our global fit we find
$$\hat{\chi}^2({\rm LMA~sterile}) = \hat{\chi}^2({\rm LMA~active}) +29.$$
SMA and LOW provide sterile fits slightly worse than LMA.

\subsubsection*{Autopsy of `2+2' oscillations}
The main motivation for adding a sterile neutrino, $\nu_{\rm s}$,
is provided by the LSND anomaly.
The extra neutrino can then be used to generate
either the solar or atmospheric or LSND oscillations
(so that active neutrinos can generate the remaining two anomalies).
Only the last possibility is now not strongly disfavoured.
In fact, taking into account that SK and MACRO atmospheric data
give a $6\sigma$ evidence for $\nu_\mu\to \nu_\tau$ versus
$\nu_\mu\to \nu_{\rm s}$~\cite{atmsterile},
it is no longer possible to 
fit the solar or atmospheric evidences (or one combination) with one sterile neutrino:
`2+2' schemes predict that
the fraction of sterile involved in solar oscillations,
plus the fraction of sterile involved in atmospheric oscillations,
adds to one~\cite{2+2},
but both solar and atmospheric data refuse sterile neutrinos.
In fig.~\ref{fig:sterile}a we show
how SK/SNO rates and solar model predictions for the Boron flux constrain
the sterile fraction in solar oscillations.
In fig.~\ref{fig:sterile}b we show how strongly
data favor $\eta_{\rm s}\approx 0$ in both
solar and atmospheric oscillations.
The `number of standard deviations' in
fig.~\ref{fig:sterile}b is computed as $N_\sigma = [\chi^2(\eta_{\rm s}) - \chi^2(\eta_{\rm s} = 0)]^{1/2}$,
where $\chi^2$ is
minimized with respect to the Boron flux.
Atmospheric bounds have been compiled from~\cite{atmsterile}
(na\"{\i}vely interpolating the few points given in~\cite{atmsterile}),
where significant but unpublished data are employed.
Solar bounds have been computed from SNO data alone
 (assuming an energy-independent $P_{ee}$),
and from a global fit of all solar data.
The result is not significantly different in the two cases, and
depends on solar model uncertainties
(e.g.\ the choice of $S_{17}$)\footnote{If LMA is the right solution,
it will be possible (assuming CPT-invariance) to put safe bounds on the sterile
fraction when KamLand will have measured the oscillation parameters
$\theta$ and $\Delta m^2$ using reactor $\bar{\nu}_e$.
This issue has recently been precisely studied~\cite{KamBGP}.
If LOW were the right solution,
solar-model-independent
bounds on the sterile fraction can be derived from
Borexino (and maybe KamLand) data about earth matter effects in Beryllium neutrinos.
In fact, the oscillation length in matter is roughly proportional to  $1/(1-\eta_{\rm s}/2)$,
assuming that the earth density is known.}.
% Having a total sterile fraction of one in solar and atmospheric oscillations
% ($\eta_{\rm s}^{\rm sun} + \eta_{\rm s}^{\rm atm}$)...

We remark that the bound on the sterile fraction
holds in the case of active/sterile mixing at a $\Delta m^2 \gg 10^{-3}\eV^2$,
as suggested by the LSND anomaly.
Active/sterile mixing with two small $\Delta m^2 $
(e.g.\ one $\Delta m^2$ around $10^{-4}\eV^2$ and the other around $10^{-11}\eV^2$)
could still give
large effects at sub-MeV neutrino energies.

\label{7out}

\paragraph{Acknowledgements} We thank Carlos Pe\~na-Garay,
Andrea Gambassi, Alberto Nicolis and Donato Nicol\`o for useful discussions. We
thank Y.~Suzuki for the unpublished SuperKamiokande data.

\appendix

\section{Details of the computation}
The energy spectra for the independent components of the solar neutrino
flux have been obtained from~\cite{BahcallWWW}.
The neutrino production has been averaged for each flux component
over the position in the sun
as predicted
in~\cite{BP98,BahcallWWW}.
This averaging does not give significant corrections.
MSW oscillations inside the sun have been taken into account in the
following way.
The $3\times 3$ density matrix $\rho_S$ for neutrinos
exiting from the sun is computed using the
Landau--Zener approximation with the level-crossing probability appropriate
for an exponential density profile~\cite{MSW,Parke}.
The density profile has been taken from~\cite{BahcallWWW} and is
quasi-exponential: small corrections to $\rho_S$ have been approximately
included.
Oscillation effects outside the sun are described by the evolution matrix
$U$, so that
at the detection point  $\rho_{E}=U\rho_S U^\dagger$.
In particular, earth regeneration effects have been computed numerically
using the mantle-core approximation for the earth density profile.
We have used the mean mantle density appropriate for each trajectory as predicted by the preliminary Earth model~\cite{PREM}.
The detection cross sections in Gallium and Chlorine experiments have been taken from~\cite{BahcallWWW}, performing
appropriate interpolations.
We have used the tree-level Standard Model expression for the
neutrino/electron cross section at SK.

The total neutrino  rates measured with the three
kinds of experimental techniques are~\cite{ClSun,KaSun,GaSun,ExpsSun}
\begin{eqnsystem}{sys:S}
R_{\rm Cl}|_{\rm exp} &=& (2.56\pm 0.22)\,{\rm SNU}\label{eq:clsig}\\
R_{\rm Ga}|_{\rm exp} &=& (74.7 \pm 5)\,{\rm SNU}
\label{eq:gasig}\\
R_{\rm SK}|_{\rm exp} &=& (2.40\pm0.08)\cdot 10^{6} \,{\rm cm}^{-2}{\rm
s}^{-1}\label{eq:sksig}
\end{eqnsystem}
where $\hbox{SNU}\equiv 10^{-36}\,\hbox{interactions per target atom and per second}$.
We have combined systematic errors in quadrature with statistical errors.
The probability distribution function $p({\rm data}|{\rm theory})$ is computed using the covariance matrix
described in~\cite{LisiChiq,recentfits1}.
%  \footnote{Although we follow this definition, we notice that
%  improving the treatment of the errors on the detection cross sections
%  by taking into account the energy dependence of the survival $\nu_e$ probability
%  could make some small difference for the SMA region.  See also~\cite{giunti0}.}.
Around the best-fits, it agrees well with the simpler pdf used in~\cite{noSSM}.
The experimental energy resolution at SK has been taken into account as suggested
in~\cite{ExpResolution,Pena}.

The solar-model-independent SK data included in the fit are
the energy spectrum of the recoil electrons measured separately at SK during  the day and during the night.
Each energy spectrum~\cite{grazieSuzuki} is composed of 17 energy bins
between $5.5$ and $14\MeV$, plus one bin between $14$ and $20\MeV$.
For these data we have used the pdf suggested by the SK collaboration~\cite{grazieSuzuki}, and
described in~\cite{Pena} for a slightly different set of data.

\medskip

The results presented in the addenda
have been obtained using updated data and a slightly improved fitting procedure,
as described in section 6 and 7.

% We have computed the pdf for all values of $\Delta m^2$ and $\theta$ on a non uniform $54\times 50$ grid 
% that covers the range $\Delta m^2 = 10^{-(10\div 3)}\eV^2$, $\tan\theta = 10^{-4\div 1}$.

% There is no reliable solar model prediction for the flux of hep neutrinos;
% the most energetic bin allows to measure them, but with a large error.
% We have omitted from the analyis such energy bin and set the hep flux to the value predicted in BP98.
% The remaining data are not significantly affected by hep neutrinos, even if their flux is
% few times larger than the BP98 central value.

\footnotesize
\begin{multicols}{2}

\end{multicols}

\newpage

\normalsize

~

\vspace{-25mm}

\section*{8~~~Addendum: first KamLAND results}\label{8in}
In this addendum, we  update our results by adding
the  \includegraphics[width=18mm]{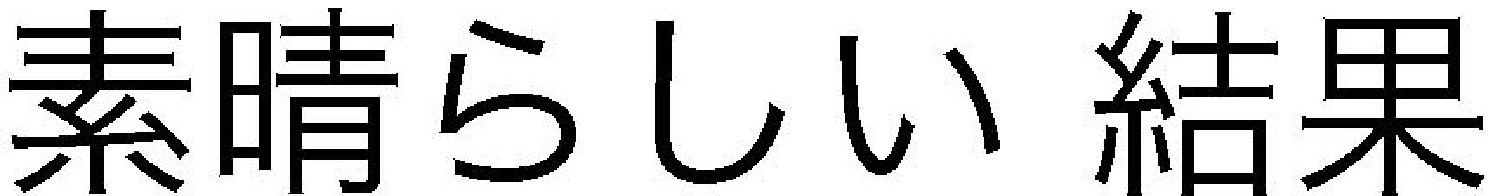} recently announced by KamLAND~\cite{KamLAND}.
We also update our previous analysis including the most recent data from GNO and the full set of spectral and day/night SNO data.
We here list the set of data included in our fit.
\begin{itemize}
\item Reactor anti-neutrino data (27 bins):
\begin{itemize}
\item The KamLAND data, divided in 13 energy bins  with prompt energy higher than $2.6\MeV$~\cite{KamLAND}.\footnote{
The energy spectrum of $\bar{\nu}_e$ emitted by a nuclear reactor
can be accurately approximated as~\cite{reactors}
$$\frac{dn}{dE} =  \frac{W}{\sum_j f_j E_j} \sum_i   f_i  \exp (a_{0i} + a_{1i} E + a_{2i} E^2)$$
where $E = E_\nu/\MeV$.
The sums run over the isotopes, $i,j=\{^{235}\hbox{U} , ~^{239}\hbox{Pu} , ~^{238}\hbox{U} , ~^{241}\hbox{Pu} \}$,
which fissions produce virtually all the total thermal power $W$.
Their relative abundancies $f_i$, and
the other numerical coefficients are~\cite{KamLAND,reactors}
$$
\begin{array}{c|cccc}
\hbox{Isotope} &
^{235}\hbox{U} & ^{239}\hbox{Pu} & ^{238}\hbox{U} & ^{241}\hbox{Pu} \cr \hline
\hbox{Relative abundancy $f_i$} &56.8\% & 29.7\% & 7.8\% & 5.7\% \cr 
\hbox{Energy per fission $E_i$ in $\MeV$ } &201.7 & 205.0 & 210.0 & 212.4 \cr 
a_0 &0.870 & 0.896 & 0.976 & 0.793 \cr 
a_1 &-0.160 &    -0.239 & -0.162 & -0.080 \cr 
a_2 & -0.091 & -0.0981 & -0.079 & -0.1085 
\end{array}$$
$\bar\nu_e$ are detected using the reaction
$\bar{\nu}_e p \to \bar{e} n$~\cite{sigma}.
Using a scintillator, reactor experiments can see both the $\gamma$ ray emitted when the neutron $n$ is captured,
and the scintillation emitted
by the positron $\bar{e}$ as it moves and finally annihilates with an $e$.
The total measured energy is $E_{\rm vis} = E_{\bar{e}}+ m_e$, and 
$E_{\bar{e}}$ is related to the neutrino energy by the kinematical relation
$E_\nu = E_{\bar{e}} + K_n +m_n -m_p$ (where $K_n$ is the small neutron kinetic energy;
A.S. thanks F. Vissani for help in including it in our numerical code
for the cross-section $\sigma$).
The energy resolution is $\sigma_E = 7.5\% \sqrt{E_{\rm vis}\cdot \MeV}$~\cite{KamLAND}.
The KamLAND scintillator contains $N_p=3.46~10^{31}$ free protons in its fiducial volume.
Summing over all reactors $r$ (that emit a power $W_r$ from a distance $d_r$),
 the number of neutrino events in any given range of $E_{\rm vis}$ is
$$ n =\sum_r
\int dE_\nu ~P_{ee}(d_r,E_\nu) 
\sigma(E_\nu) N_p \frac{dn_r/dE_\nu}{4\pi d_r^2} 
\int_{E_{\rm vis}^{\rm min}}^{E_{\rm vis}^{\rm max}} dE_{\rm vis}   
\frac{e^{-(E_{\rm vis} - E_\nu + K_n+0.782\MeV)^2/2\sigma_E^2}}{\sqrt{2\pi}\sigma_E} 
$$
Neglecting small differences between the energy spectra of different reactors
(i.e.\ assuming that all reactors have the same relative isotope abundancies $f_i$),
the spectrum of detected neutrinos $n$ can be conveniently rewritten as
$
n=n_0 \langle P_{ee}\rangle $
where $n_0$ is the rate expected in absence of oscillations
and $\langle P_{ee}\rangle$ is the survival probability,
appropriately averaged over energy and distance.
% $$\langle P_{ee}\rangle =\sum_r p_r P_{ee}(E_\nu, d_r)\qquad
% p_r = \frac{W_r/d_r^2}{\sum_{r'}  W_{r'}/d_{r'}^2}$$
% and $R$ is the response function.
Assuming an energy-independent suppression, we find $P_{ee}=0.625\pm 0.094$,
in agreement with the more accurate KamLAND result.
For recent related papers see~\cite{recenti}.}

\item The CHOOZ data, divided in 14 bins and fitted as in analysis ``A'' of~\cite{CHOOZlast}.

\end{itemize}
\item Solar neutrino data (80 bins):
 \begin{itemize}

\item The SNO data~\cite{SNOlast}, divided in 34 bins (17 energy bins times 2 day/night bins).\footnote{SNO rates 
receive contributions from CC, NC, ES and background events.
SNO extracts from all data (energy and zenith-angle spectra) 
the CC, NC and ES components.
We instead extract the CC, NC components from the energy spectrum,
assuming the standard relation $\Phi_{\rm ES} \approx \Phi_{\rm CC} + 0.15 \Phi_{\rm NC}$
(the ES rate has also been accurately measured by SK).
Assuming energy-independent oscillations between active neutrinos,
our reanalysis of SNO data gives
$$
\Phi_{\rm NC} =\Phi_{^8\rm B}  = (5.2 \pm 0.5)~10^6/{\rm cm}^2{\rm s},\qquad
\Phi_{\rm CC} =  P_{ee}\Phi_{^8\rm B} = (1.76 \pm 0.08)~10^6/{\rm cm}^2{\rm s}.
$$
only slightly different from the corresponding SNO analysis~\cite{SNOlast}
(the errors are somewhat anti-correlated). }

\item The SuperKamiokande data~\cite{SKlast}, divided
in 44  zenith-angle and energy bins.

\item The Gallium rate~\cite{Galliumlast} $R_{\rm Ga} = (70.4 \pm 4.4)\,{\rm SNU}$, obtained averaging
the most recent  SAGE, Gallex and GNO data.

\item The Chlorine rate~\cite{Chlorinelast}
$R_{\rm Cl} = (2.56 \pm 0.23)\,{\rm SNU}$.

\end{itemize}
% Experimental uncertainties are taken into account as described by the collaborations.

\item Solar model predictions.
We assume the BP00 solar model~\cite{BP00last} and its estimates of uncertainties.\footnote{In particular, we
take into account the uncertainty on the energy spectrum of $^8$B neutrinos, that affects in a correlated way SK, SNO and the other solar experiments.}
% Survival probabilities in the sun are computed using the 
% accurate approximations from~\cite{MSWlast}.

\item Earth matter effects are computed in the 
mantle/core approximation, improved by
using the average density appropriate for each 
trajectory as predicted in~\cite{PREMlast}.

\end{itemize}
We fit these data assuming $\nu_{e}\leftrightarrow \nu_{\mu,\tau}$
 neutrino oscillations with a single $\Delta m^2$ and a mixing angle $\theta$.
The total $\chi^2$ is obtained by summing the contribution from solar $\nu$ data,
$\chi^2_{\nu}$, with the contribution from reactor $\bar\nu$ data,
$ \chi^2_{\bar{\nu}} = -2\ln {\cal L}$
(the likelihood ${\cal L}$ is computed employing when appropriate the Poissonian distribution).
Best-fit regions are computed in Gaussian approximation,
cutting the $\Delta\chi^2$ at the value appropriate for 2 degrees of freedom.

\begin{figure}
\vspace{-25mm}
$$\hspace{-6mm}\includegraphics[width=6cm]{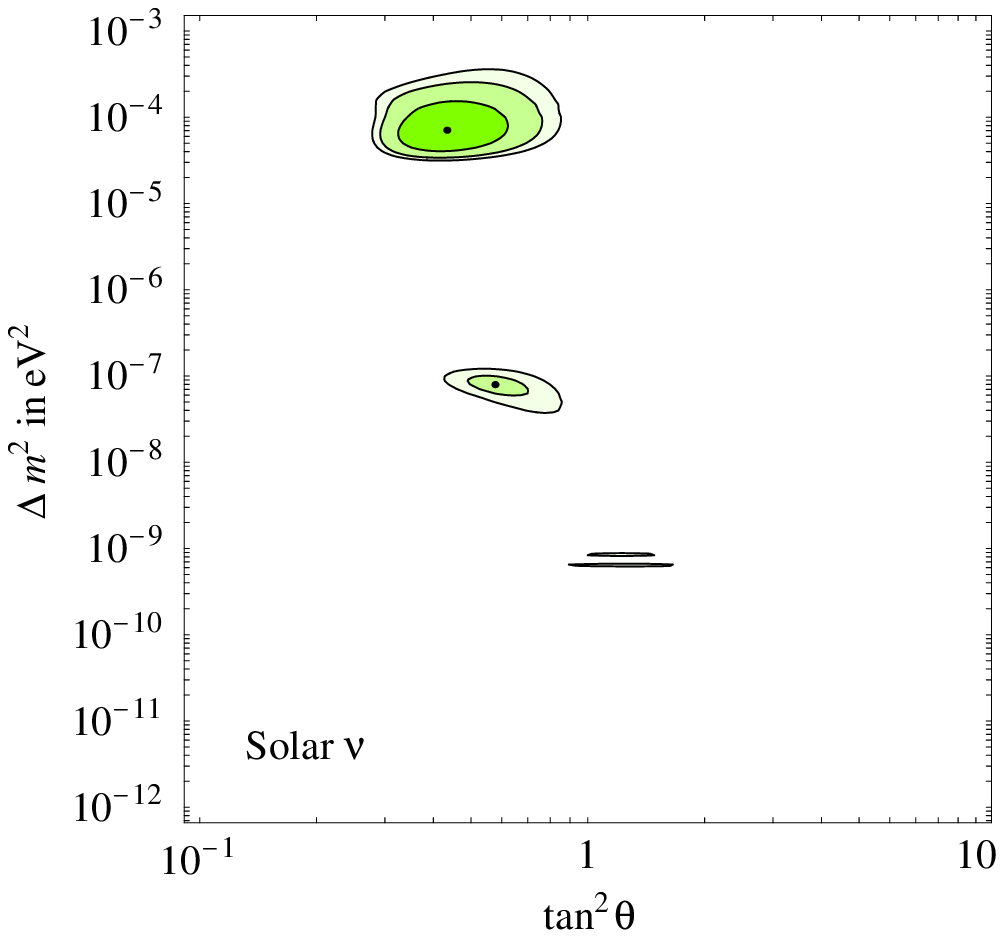}\includegraphics[width=6cm]{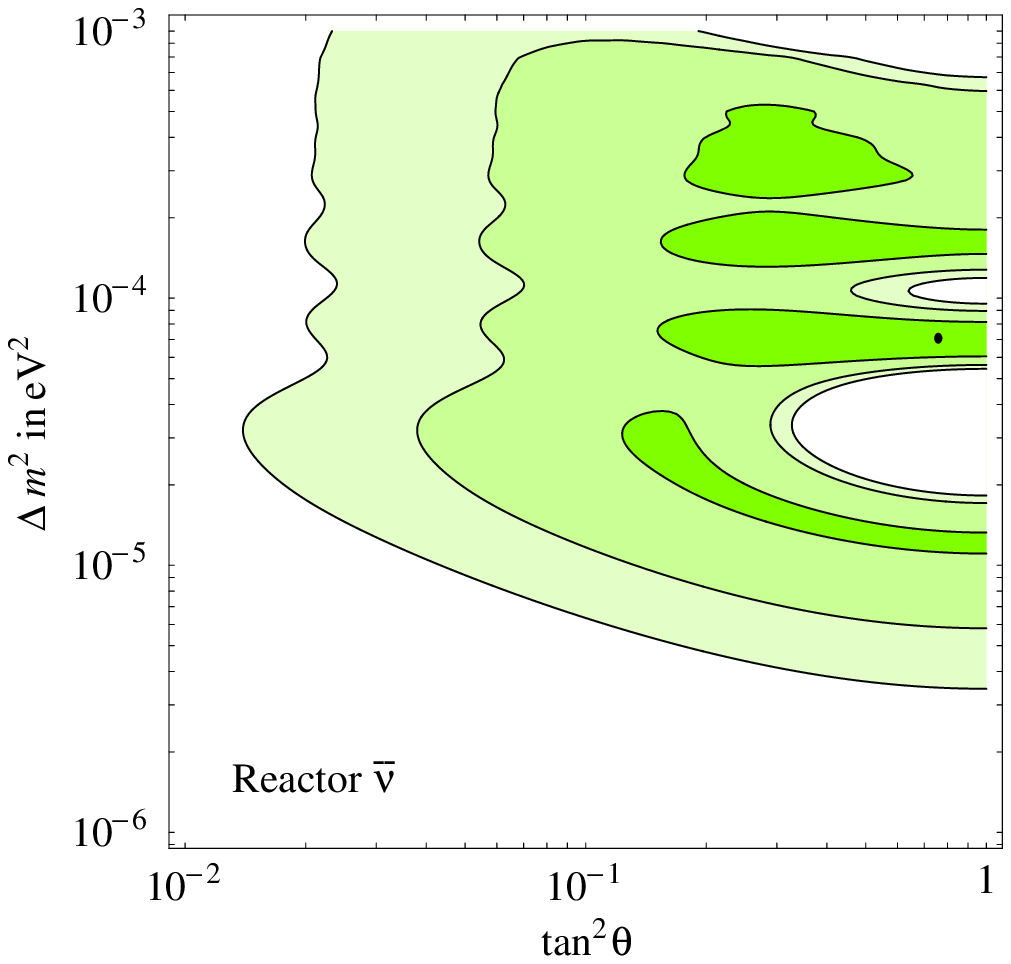}\includegraphics[width=6cm]{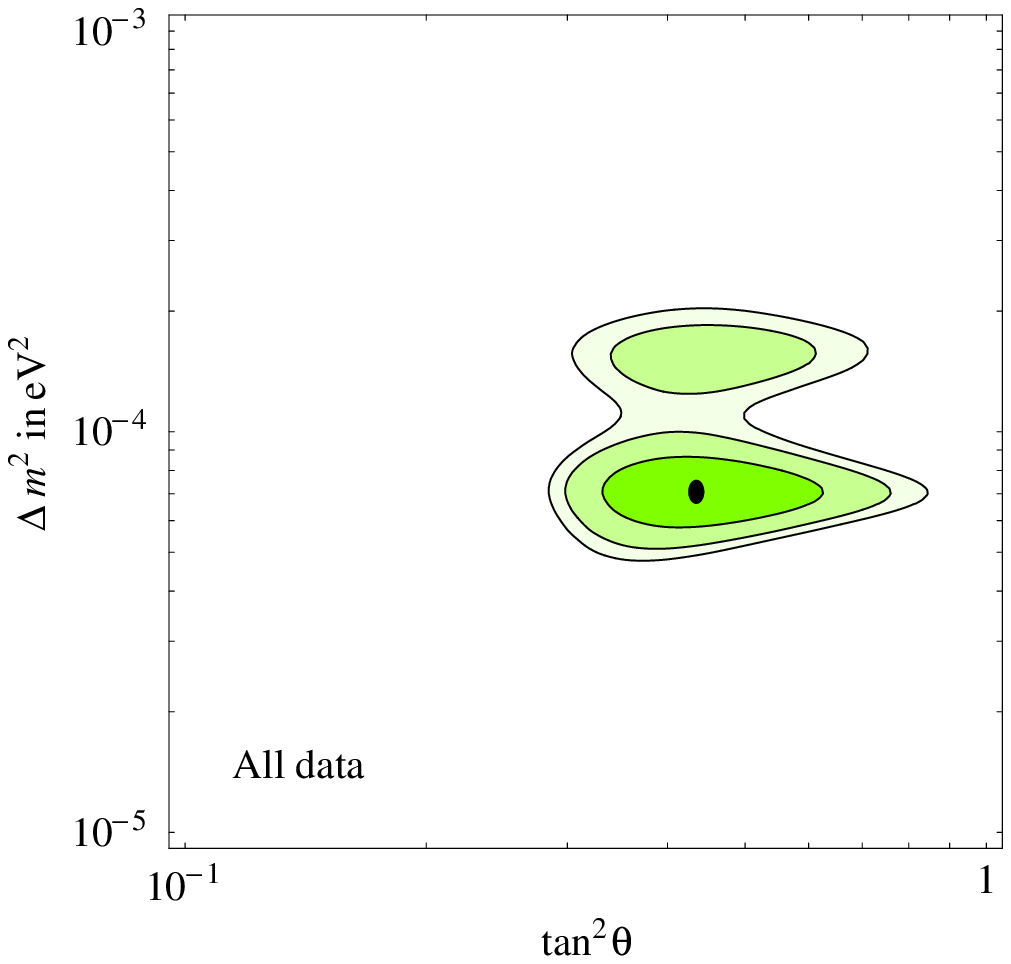}$$
\vspace{-5mm}
\caption{\label{fig:KL}\em Best-fit regions at $90,~99$ and $99.73\%$ {\rm CL} obtained fitting
(a) solar $\nu$ data;
(b) reactor $\bar{\nu}$ data (that do not distinguish $\theta$ from $\pi/2-\theta$);
(c) all data (we find no other allowed regions at the considered CL outside the plotted range).}
\end{figure}

% The result of our fit is shown in fig.~\ref{fig:KL}.
% The best fit point (marked with a dot) is at
% $$\Delta m^2 = 7.0~10^{-5}\eV^2,\qquad
% \tan^2\theta = 0.45 \qquad\hbox{where}\qquad
% \chi^2_{\nu} = 72.9.$$
% % (for reactors Poisson...)
% The $99\%$ CL ranges for the single parameters (1 dof) are
% $$5.5~10^{-5}\eV^2<\Delta m^2 <1.0~10^{-4}\eV^2,\qquad
% 0.32<\tan^2\theta < 0.77.$$
The result of our fit is shown in fig.~\ref{fig:KL}.
The best fit point (marked with a dot) and the $1\sigma$ ranges for the single parameters (1 dof) are
$$\Delta m^2_{12} = (7.0 \pm 0.6)\,10^{-5}\eV^2,\qquad
\tan^2\theta_{12} = 0.45 \pm  0.06.$$
% (for reactors Poisson...)
At the best fit point $\chi^2_{\nu} = 72.9$ (80 bins) and $\chi^2_{\rm KamLAND}\approx 6$ (13 bins\footnote{We have normalized the likelihood such that ${\cal L} =1$ 
when data agree with theorethical predictions, and
ignored all other issues related to the difference between Poissonian and Gaussian distributions.}). 
The disfavoured local minimum 
at higher $\Delta m^2 = 1.53~10^{-4}\eV^2$ shown in fig.~\ref{fig:KL}c (`HLMA' solution)
has a $\chi^2$ 6.4 higher than in the best fit solution.
According to our global fit, the combined evidence for a solar neutrino anomaly is almost $10\sigma$;
the evidence for LMA versus SMA is about $6\sigma$;
the QVO  and LOW solutions are disfavoured at about $(4\div 5)\sigma$.
According to a na\"{\i}ve Pearson $\chi^2$ test these excluded solutions still have an acceptable goodness-of-fit, around $20\%$
(this happens because the $\chi^2$ test has a limited statistical power when many data are fitted together, as we explained in section~4)
while our more efficient test recognizes that their goodness-of-fit probability is $\sim 10^{-3}$.
More importantly, present data are well described by LMA oscillations and do not contain hints for something more.

% a correct estimate is
% % $$\hbox{GOF(LMA)}\sim 50\%,\qquad
% % \hbox{GOF(LOW, QVO)}\sim 0.1\%,\qquad\hbox{GOF(SMA)}\sim 0.$$

KamLAND data have an important impact also  on various related issues:
\begin{itemize}
\item {\bf Maximal mixing?}
Fitting reactor data alone (fig.~\ref{fig:KL}b), we find the best fit point at 
$\Delta m^2 = 6.9~10^{-5}\eV^2$,
$\sin^2 2\theta = 0.99$
but also a larger uncertainty on the mixing angle $\theta$.
Therefore maximal mixing remains disfavoured by solar data.
In the global fit of fig.~\ref{fig:KL}c, maximal mixing 
would enter the best-fit regions only at $99.96\%$ CL (2 dof).
Before KamLAND, maximal mixing was allowed in
Q(VO) solutions at a reasonable CL.

\item {\bf Oscillations?}
If the present trend is confirmed by future KamLAND data,
all proposed alternatives to oscillations will be excluded.
If $\Delta m^2 \circa{<}2~10^{-4}\eV^2$,
with more statistics KamLAND can see a clean oscillation pattern, giving
a precise measurement of $\Delta m^2$~\cite{noSSMlast}.
This will tell if earth-matter effects can be observed by a larger SK-like solar detector.
At the best-fit point the total day/night asymmetry is predicted to be
$2.7\%$ at SK and $4.5\%$ in CC events at SNO,
comparable to present sensitivities.

\item {\bf Future experiments.}
According to the present oscillation best-fit point,
Borexino should observe no day/night asymmetry, no seasonal variation
and a $65\%$ reduction of the $^7$Be flux.
If this will happen, solar experiments alone will tell that
LMA is the unique oscillation solution,
confirming the CPT theorem.
The impact  of KamLAND data on our knowledge of solar neutrinos will be discussed in an update to~\cite{noSSMlast}.
Knowing that LMA oscillations are the solution to the solar neutrino anomaly
allows to plan which  experiment(s) will give the most precise measurement of $\Delta m^2$ and $\theta$
(see~\cite{subMeV} and a future update).

% Future solar neutrino experiments might compete with KamLAND in giving a precise 
% determination of $\theta$.

\item {\bf Three flavours}.
When analyzed in a three-neutrino context,
solar and reactor experiments are also affected by atmospheric oscillations if $\theta_{13}\neq 0$.
For the observed value of $\Delta m^2_{\rm atm}$, CHOOZ gives the dominant upper bound on $\theta_{13}$.
Therefore setting $\theta_{13}=0$ is an excellent approximation for fitting solar data.

\item {\bf Neutrino-less double-beta decay}.
Assuming hierarchical neutrinos $m_1\ll m_2 \ll m_3$
the $ee$ element of the neutrino mass matrix probed by
$0\nu2\beta$ decay experiments
can be written as $|m_{ee}| = |m_{ee}^{\rm sun} + e^{i \alpha} m_{ee}^{\rm atm}|$
where $\alpha$ is an unknown Majorana phase and 
the `solar' and `atmospheric' contributions can be predicted from oscillation data.
Including KamLAND data, the solar contribution to $|m_{ee}|$
is $m_{ee}^{\rm sun} =( 2.6\pm 0.4)~\,\hbox{meV}$.
The CHOOZ bound on $\theta_{13}$ implies $m_{ee}^{\rm atm} < 2\,\hbox{meV}$ at 99\% CL.

\item {\bf Sterile?}
Experimental bounds on a possible sterile neutrino component involved in solar oscillations are not yet significantly affected
by KamLAND data, that instead have a strong impact
on bounds from big-bang nucleosynthesis~\cite{bbn}.
Such bounds cannot be avoided by modifying cosmology before neutrino decoupling
and are usually presented in the $(\theta, \Delta m^2)$ plane assuming pure sterile oscillations.
It is now more interesting to know the bound on a possible subdominant sterile component involved in LMA solar oscillations.
We estimate that this bound is competitive with the direct experimental bound.

\item {\bf LSND.}
Solutions to the LSND anomaly based on extra sterile neutrinos are now disfavoured by other experiments~\cite{CPT}.
The CPT-violating neutrino spectrum proposed in~\cite{CPTth} allowed to fit all neutrino data,
but predicted no effect in KamLAND.
A slightly different CPT-violating solution mentioned in~\cite{CPT} can give a suppression in KamLAND at the price
of an up/down asymmetry of atmospheric muon neutrinos $A\approx 20\%$, somewhat smaller than the
measured value.
This possibility is more disfavoured than `3+1 oscillations' and less disfavoured than `2+2 oscillations',
as will be discussed in an update to~\cite{CPT}.

\end{itemize}

\label{8out}

\footnotesize
\begin{multicols}{2}
 
\end{multicols}

\newpage

\normalsize

\begin{figure}
\vspace{-25mm}
$$\hspace{-6mm}\includegraphics[width=6cm]{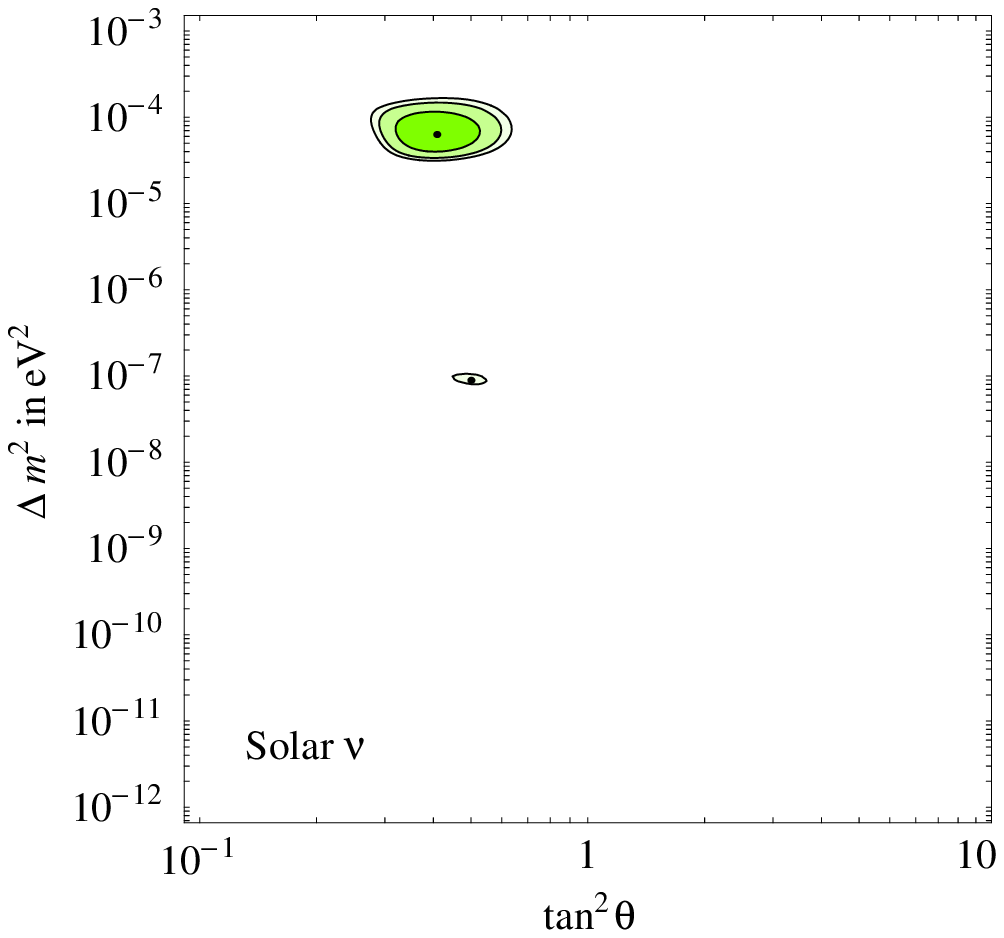}\includegraphics[width=6cm]{reactorOnly}\includegraphics[width=6cm]{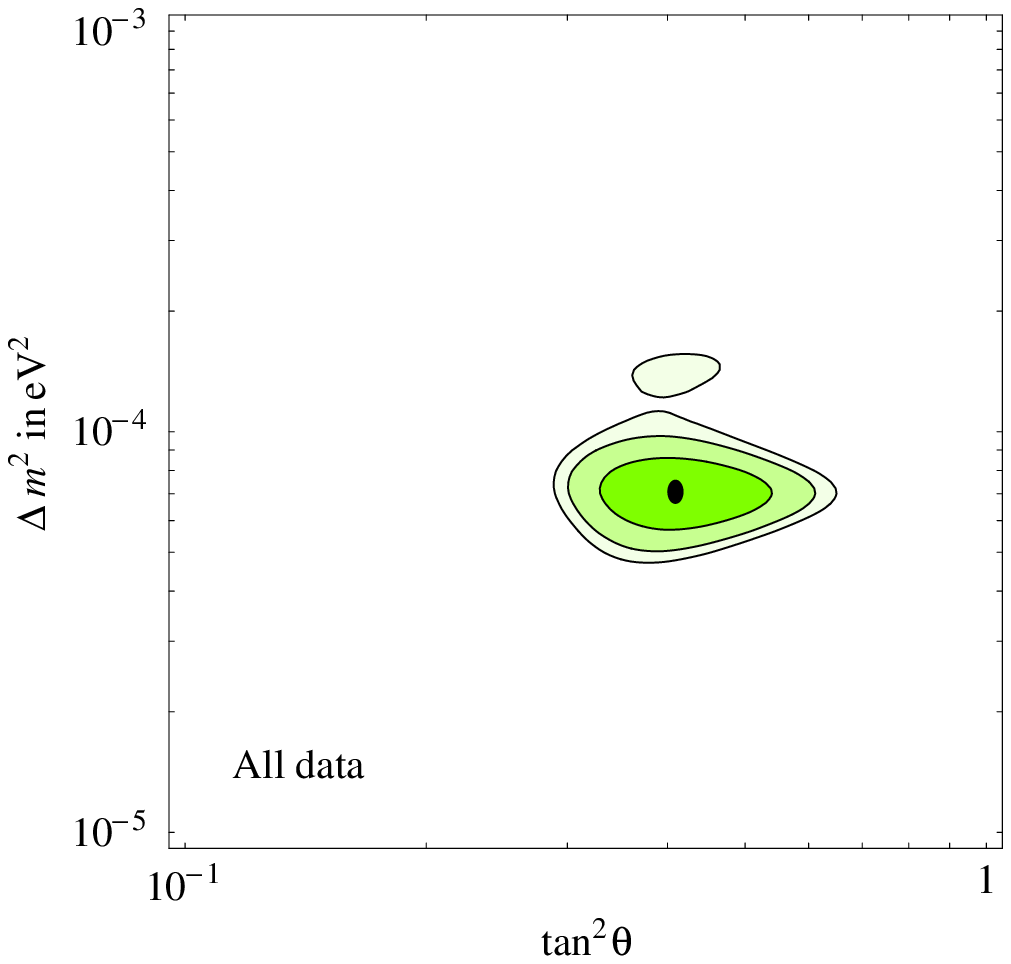}$$
\vspace{-5mm}
\caption[x]{\label{fig:SNOsalt}\em 
Best-fit regions at $90,~99$ and $99.73\%$ {\rm CL} obtained fitting
(a) solar $\nu$ data;
(b) reactor $\bar{\nu}$ data (that do not distinguish $\theta$ from $\pi/2-\theta$);
(c) all data.
These fits can be directly compared with the previous ones, shown in fig.~\ref{fig:KL}.}
\end{figure}

\section*{9~~~Addendum: SNO data with enhanced NC sensitivity}\label{9in}
SK detects solar $\nu$ using water as target.
SNO now employs heavy salt water~\cite{SNOsalt}.
Heavy water allowed to study the NC reaction $\nu_{e,\mu,\tau} d\to \nu_{e,\mu,\tau} np$.
Salt allows to tag the $n$ with higher efficiency
(producing multiple $\gamma$ rays from neutron capture)
and therefore a more accurate measurement of total solar $\nu_{e,\mu,\tau}$ flux.
However, the enhanced NC signal makes more difficult to study the CC energy spectrum.
As a consequence the recently released `salty' SNO data~\cite{SNOsalt} can be conveniently summarized
as a measurement of the total CC, NC and electron scattering (ES) neutrino fluxes.
Uncertainties on the Boron flux, its energy spectrum, and cross sections are correlated with other experiments.
We fit latest SNO data as suggested by the SNO collaboration
(this means that we are slightly less accurate than the SNO collaboration itself).
We employ the most recent value of the Gallium rate~\cite{Ga2003}.

The result of our fit is shown in fig.~\ref{fig:SNOsalt}, where
the best fit point is marked with a dot.
The $1\sigma$ ranges for the single parameters (1 dof) are
$$\Delta m^2_{12} = (7.2 \pm 0.7)\,10^{-5}\eV^2,\qquad
\tan^2\theta_{12} = 0.44 \pm  0.05.$$
% (for reactors Poisson...)
At the best fit point $\chi^2_{\rm solar} = 75.0$ (83 bins) and $\chi^2_{\rm KamLAND}\approx 6$.
The total evidence for a solar anomaly is $12\sigma$ (dominated by solar experiments).
We now discuss some consequences of latest SNO data:

\begin{itemize}
\item{\bf Oscillations?}
LMA solutions with larger $\Delta m^2$ (present in our previous fit of fig.~\ref{fig:KL}c)
have been disfavoured, because they predict a too large $\nu_e$ survival probability.
This is important because only if $\Delta m^2\circa{<} 2~10^{-4}\eV^2$
KamLAND should be able, with more statistics,
to see in the $\bar\nu_e$ energy spectrum
the typical spectral distortion produced by oscillations
and to measure $\Delta m^2$ accurately~\cite{noSSM3}.
So far no experiment has been able of seeing such a direct signal of
neutrino oscillations.

\item{\bf Maximal mixing} is disfavoured at $5.7\sigma$ (i.e.\ its $\chi^2$ is $5.7^2$ higher than the best LMA fit).

\item{\bf Symmetry?}
 $\tan^2\theta=1/2$ (which together with maximal $\nu_\mu\to\nu_\tau$ atmospheric
oscillations gives a  highly symmetric neutrino flavour composition,
with $|\langle\nu_{e,\mu,\tau}|\nu_2\rangle|^2 = 1/3$) 
is $\sim1.2\sigma$ above the best fit value.
Hopefully, future data will test this special value.

\item{\bf LMA}.
Omitting KamLAND $\bar\nu$ data and fitting only solar $\nu$ experiments,
the LOW solution is now disfavoured at $3.3\sigma$: $0.5\sigma$ more than without SNO salt data.
%($\chi^2_{\rm LOW} = \chi^2_{\rm LMA} + 11.2$)
Other  non-LMA oscillation solutions have now been excluded by solar data only.

\item{\bf MSW effect}. SNO gives a  $5\sigma$ evidence for
 a $\nu_e$ survival probability less than $1/2$.
 This can be explained by MSW enhancement of neutrino oscillations~\cite{MSW3},
 while averaged vacuum oscillations of two-neutrinos
can only give  $P_{ee}=1-\frac{1}{2}\sin^22\theta\ge 1/2$
(see also~\cite{LisiMSW}).
Three-neutrino mixing can only marginally reduce $P_{ee}$,
since SK and SNO imply that $\theta_{13}$ is small.

\item{\bf Three flavours}.
When analyzed in a three-neutrino context,
solar and reactor experiments are also affected by atmospheric oscillations if $\theta_{13}\neq 0$.
For the observed value of $\Delta m^2_{\rm atm}$, CHOOZ gives the dominant upper bound on $\theta_{13}$.
$\theta_{13}=0$ remains an excellent approximation for fitting solar data
even taking into account that a recent SK analysis slightly reduces $\Delta m^2_{\rm atm}$,
thereby enlarging the combined CHOOZ and SK range of
$\theta_{13}$
from $\sin^22\theta_{13} = 0\pm0.065$~\cite{FSV}
to $0\pm0.085$. (See also~\cite{LisiSK'}).

\item {\bf Neutrino-less double-beta decay}
experiments probe the $ee$ element of the neutrino mass matrix.
Newer SNO data (and the revised bounds on $\theta_{13}$) slightly
shift the range of  $|m_{ee}|$ compatible with oscillations~\cite{FSV}.
Assuming hierarchical neutrinos $m_1\ll m_2 \ll m_3$ and proceeding as in~\cite{FSV} we get
$$
|m_{ee}|=(0.8\div 4.2) \meV\hbox{ at $90\%$ CL}\qquad\hbox{ and }\qquad
|m_{ee}|=(0\div 5.3) \meV \hbox{ at $99\%$ CL.}
$$
In the case of  inverted hierarchy (i.e.\ $m_3\ll m_1\approx m_2$, or $\Delta m^2_{23}<0$) 
the updated range is
$$
|m_{ee}|=(15\div 57) \meV\hbox{ at $90\%$ CL}\qquad\hbox{ and }\qquad
|m_{ee}|=(12 \div 61) \meV \hbox{ at $99\%$ CL.}
$$
Latest SNO data imply a slightly stronger bound 
on the mass $m_\nu$ of quasi-degenerate neutrinos~\cite{FSV}:
$$
m_{\nu}< 0.9 ~(1.2)\, h\,  \eV\qquad\hbox{ at 90 ($99\%$) CL.}
$$
This bound is based on the fact that a
 large $m_{\nu}$ is compatible with the Heidelberg-Moscow upper bound on $|m_{ee}|$~\cite{HM3} 
only if  $\theta_{12}\approx \pi/4$.
The factor $h\approx 1$, precisely defined in~\cite{FSV},
 parameterizes the uncertainty in the 
$0\nu2\beta$ $^{76}$Ge nuclear matrix element.

\item{\bf Robustness}.
A few years ago global analyses were the only way of
extracting the main features of the solar neutrino anomaly from the data.
These analyses required a lot of work:
a careful simulation of the sun, 
of the neutrino flavour evolution in the sun, in vacuum and in the earth,
a careful estimation of all possible sources of systematic and theorethical errors.
Global fits favoured the LMA solution before KamLAND confirmed it.

Present fits are dominated by recent data.
Their interpretation is based on few key inputs:
1) the energy spectrum of Boron neutrinos, which essentially follows from kinematics.
2) matter corrections to neutrino propagation in the sun,
which reduce the $\nu_e$ survival probability by a ${\cal O}(1)$ factor:
from $P_{ee}=1-\frac{1}{2}\sin^22\theta$ at $E_\nu\circa{<}\MeV$  (as in averaged vacuum oscillations)
to $P_{ee}\approx\sin^2\theta$
(adiabatic MSW resonance) at energies probed by SK and SNO.
Our results are  obtained  by running a precise code, but
a good approximation could be obtained 
using a simplified code.
%	\footnote{
%	Although for brevity we do not discuss its details,
%	we point out that almost all other global analyses agreed with our previous update
%	about KamLAND data, see www.cern.ch/astrumia/InstantPaper.html.}

\end{itemize}

\footnotesize
\begin{multicols}{2}
 
\end{multicols}
\label{9out}

\normalsize

\begin{figure}
\vspace{-25mm}
$$
\includegraphics[width=6cm]{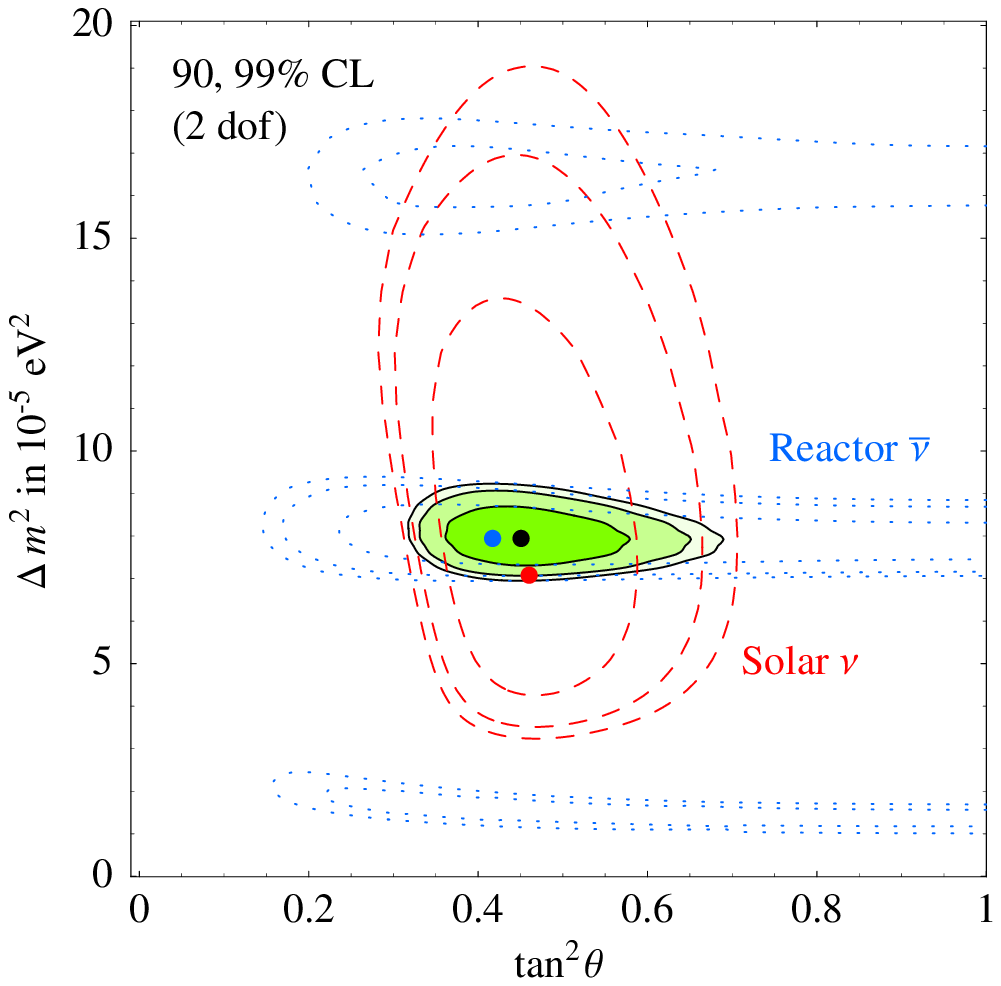}\qquad \qquad\qquad
\includegraphics[width=6cm]{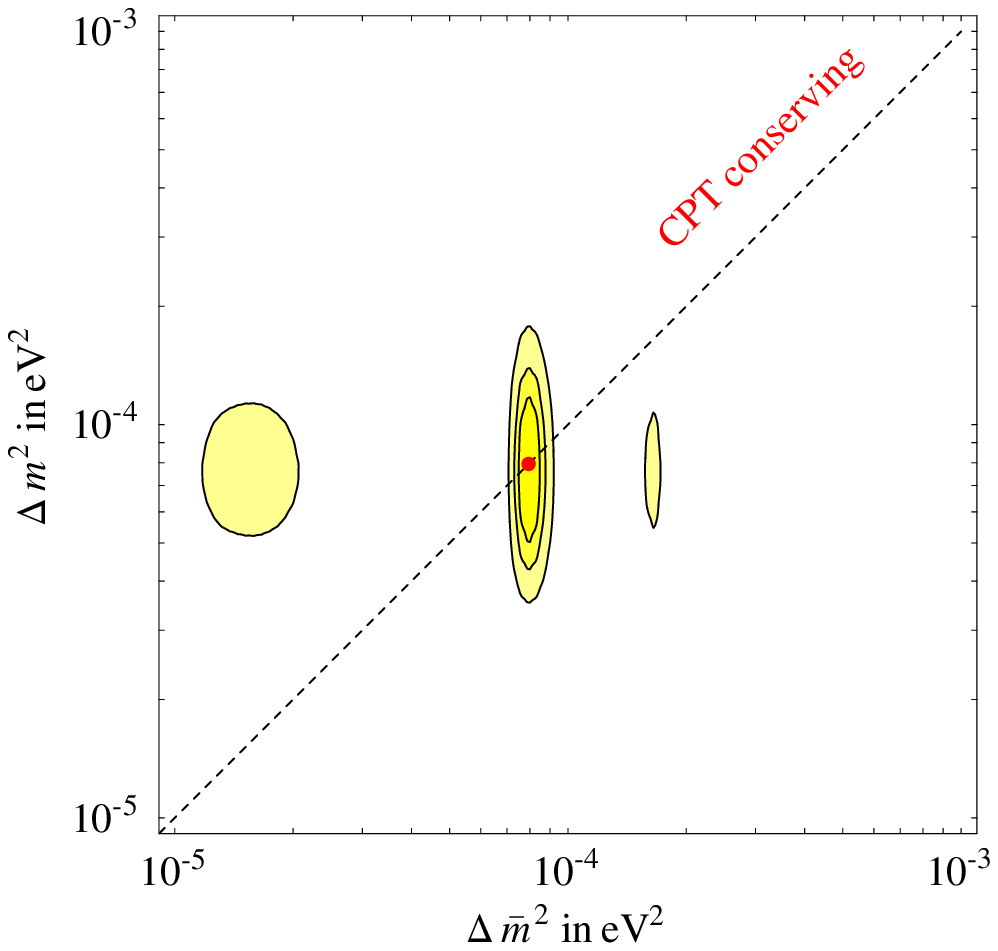}$$
\vspace{-5mm}
\caption[x]{\label{fig:2005}\em 
Best-fit regions at $90,~99$ and $99.73\%$ {\rm CL}.
Fig.~\ref{fig:2005}a assumes CPT invariance and combines
solar $\nu$ data (dashed red contours) with reactor $\bar{\nu}$ data (dotted blue contours).
The result can be directly compared with the previous one, shown in fig.~\ref{fig:SNOsalt}.
Fig.~\ref{fig:2005}b shows the separate fit for $\Delta m^2$ in neutrinos and $\Delta\bar{m}^2$ in
anti-neutrinos, marginalized with respect to mixing angles.
}
\end{figure}

\section*{10~~~Addendum: final SNO `salt' data and KamLAND}\label{10in}
We update our previous results including:
%\begin{enumerate}
the final SNO `salt' phase results~\cite{SNOsaltfinal};
KamLAND data after 766.3 ton$\cdot$yr exposure~\cite{KL2004};
the final Gallex result and the latest SAGE result~\cite{nu04}.
%\end{enumerate}
The result of our oscillation fit is shown in fig.~\ref{fig:2005}, where
the best fit point is marked with a dot.
The $1\sigma$ ranges for the single parameters (1 dof) are
\begin{equation}\label{eq:2005}
\Delta m^2_{12} = (8.0 \pm 0.3)\,10^{-5}\eV^2,\qquad
\tan^2\theta_{12} = 0.45 \pm  0.05.
\end{equation}
The total evidence for an effect
is now about 12$\sigma$ in solar data and about 6$\sigma$ in KamLAND data.

We remark that
our results are based on a careful global fit,
that was necessary to interpret the data available a few years ago.
But present data speak by themselves, and a simple approximate
analysis is sufficient to get results practically equivalent to the full ones in eq.\eq{2005}.
Indeed $\Delta m^2_{12}$ is directly determined by the position of the oscillation dip at
KamLAND, with negligible
contribution from all other experiments.\footnote{This will be rigorously true in the future.
For the moment solar data are needed to eliminate spurious solutions
mildly disfavored by KamLAND data, as illustrated in fig.~\ref{fig:2005}a.}
The mixing angle is directly determined by the SNO measurements\footnote{Taking into account the SK measurement
$\Phi(\nu_{e})  + 0.155 \Phi(\nu_{e,\mu,\tau}) = (2.35\pm0.06)~10^6/\cm^2\sec $
and the solar model prediction
$\Phi(\nu_{e,\mu,\tau}) = (5.05\pm 0.86)~10^6/\cm^2\sec $
would only marginally improve the measurement of $\langle P_{ee}\rangle$.}
$$\langle P_{ee} \rangle \equiv  {\Phi(\nu_{e}) }/{\Phi(\nu_{e,\mu,\tau}) }=0.35\pm0.03.$$
%(we neglected small statistical correlations between NC and CC data,
%and summed errors in quadrature).
In the adiabatic limit LMA predicts $\langle P_{ee}\rangle \simeq \sin^2\theta$.
Well known analytic formul\ae{} allow to take non-adiabaticity into account,
obtaining in the energy range explored by SNO:
$$\langle P_{ee} \rangle \approx 1.15\sin^2\theta_{12}
\qquad\hbox{so that}\qquad
 \tan^2\theta_{12}=0.44\pm0.06$$ 
 which agrees with the results of the global analysis in eq.\eq{2005},
 both in the central value and in its uncertainty.
 Global fits remain still useful for testing if the pieces of data not included
in our simplified analysis contain statistically significant indications for
 new physics beyond LMA oscillations.
 At the moment the answer is no.
 E.g.\ fig.~\ref{fig:2005}b shows the update of the CPT-violating solar fit of~\cite{CPT}.
To really conclude, the experimental results
auspicated in the conclusions of our first version
have been achieved, 
removing the motivation for our analyses.

\footnotesize
\begin{multicols}{2}
 
\end{multicols}
\label{10out}

\end{document}